\documentclass[sigconf,nonacm,timestamp]{acmart}

\AtBeginDocument{%
  }

\setcopyright{acmlicensed}
\copyrightyear{2025}
\acmYear{2025}
\acmDOI{XXXXXXX.XXXXXXX}

\acmConference[ACM CUI 2025]{ACM CUI 2025}{July 8,
  2025}{Waterloo, Ontario, Canada}
\acmISBN{978-1-4503-XXXX-X/18/06}

\usepackage{tabularx}
\usepackage{adjustbox}
\usepackage{geometry}
\usepackage{enumitem}
\usepackage{tcolorbox}
\usepackage{amsmath}
\usepackage{multicol}
\usepackage{svg}
\usepackage{tikz, pgfplots}
\pgfplotsset{compat=1.18}
\usepackage{subcaption}
\usetikzlibrary{matrix}
\usepackage{xcolor}
\definecolor{gray(x11gray)}{rgb}{0.75, 0.75, 0.75}

\begin{document}


\title{Static Vs. Agentic Game Master AI for Facilitating Solo Role-Playing Experiences}



\author{Nicolai H. J{\o}rgensen}
\affiliation{%
  \institution{Department of Computer Science, Aalborg University}
  \city{Aalborg}
  \country{Denmark}
}
\email{njarge20@student.aau.dk}

\author{Sarmilan Tharmabalan}
\affiliation{%
  \institution{Department of Computer Science, Aalborg University}
  \city{Aalborg}
  \country{Denmark}
}
\email{stharm20@student.aau.dk}

\author{Ilhan Aslan}
\affiliation{%
  \institution{Department of Computer Science, Aalborg University}
  \city{Aalborg}
  \country{Denmark}
}
\email{ilas@cs.aau.dk}

\author{Nicolai Brodersen Hansen}
\affiliation{%
  \institution{Department of Computer Science, Aalborg University}
  \city{Aalborg}
  \country{Denmark}
}
\email{nbha@cs.aau.dk}

\author{Timothy Merritt}
\affiliation{%
  \institution{Department of Computer Science, Aalborg University}
  \city{Aalborg}
  \country{Denmark}
}
\email{merritt@cs.aau.dk}


\renewcommand{\shortauthors}{J{\o}rgensen et al.}

\begin{abstract}
This paper presents a game master AI for single-player role-playing games. The AI is designed to deliver interactive text-based narratives and experiences typically associated with multiplayer tabletop games like Dungeons \& Dragons. We report on the design process and the series of experiments to improve the functionality and experience design, resulting in two functional versions of the system.
While v1 of our system uses simplified prompt engineering, v2 leverages a multi-agent architecture and the ReAct framework to include reasoning and action. A comparative evaluation demonstrates that v2 as an agentic system maintains play while significantly improving modularity and game experience, including immersion and curiosity. Our findings contribute to the evolution of AI-driven interactive fiction, highlighting new avenues for enhancing solo role-playing experiences.

\end{abstract}



\begin{teaserfigure}
  \includegraphics[width=\textwidth]{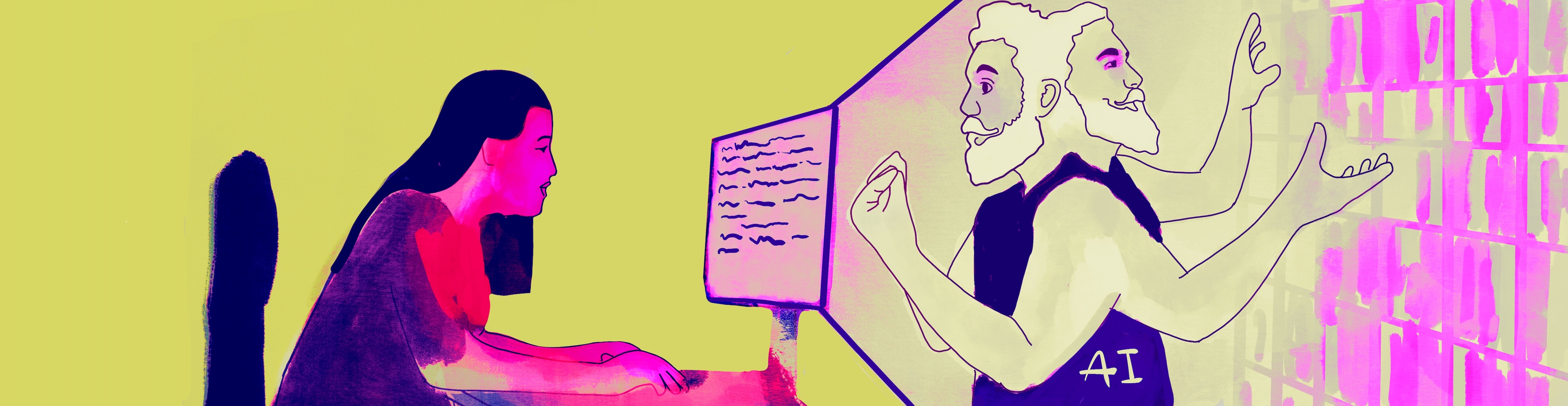}
  \caption{Conceptual image of the vision behind ChatRPG. The player at the left uses the text-based chat user interface of the role-playing game to explore and respond to the game events and take actions in the game. The Game Master AI (GM AI) builds an engaging narrative, communicates details about the fictional world, and manages the assets, events, and status to maintain coherence.}
  \Description{A conceptual image showing the vision behind ChatRPG. On the left a player is shown looking at a screen and typing text. Behind the screen is the Game Master AI depicted. One aspect of the Game Master is narrating the story to the player, while another aspect is retrieving and storing information from a shelf of books.}
  \label{fig:teaser}
\end{teaserfigure}


\maketitle

\section{Introduction} \label{sec:introduction}




In a society that is socially accelerated~\cite{rosa2019resonance}, where making deeper connections and friendships is becoming ever more challenging, designers are increasingly aiming to design for resonating experiences~\cite{piHearts_2020, curran2023social, Tan02012025}, reducing loneliness and social isolation~\cite{Baecker_et_al_2014, probst2024_et_al}, and to foster wellbeing and human flourishing~\cite{aslan2023compliment,calvo2014positive}. While loneliness and isolation are often associated with older adults~\cite{Stuart_et_al_2022}, a lesson from the pandemic is that social isolation is an important matter for all ages. Previous research has outlined various psychological benefits of conversational interactions
~\cite{clark1993stress} such as coping with stress. There is mounting evidence that role-playing games can be beneficial by engaging players in conversational interactions, even those moderated by a computer whether they involve multiple or single players~\cite{arenas_therapeutic_2022}.

Large language models (LLM), which are, in essence, conversational agents, are increasingly explored in the context of games~\cite{penny_survey_CUI2024}; a recent survey of LLM use in games is provided by Gallotta et al.~\cite{gallotta2024large}, which covers LLMs assisting human game masters and, in sporadic cases, take the role of a Game Master (GM)~\cite{hua2020playing, You_et_al_2024,Triyason2023}, which is necessary for solo role-playing. Some open-source tools already exist, like, for instance, TavernAI / SillyTavern \cite{sillytavern}, which can enhance LLM-based roleplaying with user-friendly interfaces. Similarly, there is a range of "LoRA" ("Low-Rank Adaptations" small adapter-like layers that colour the use of a specific model for a specific type of roleplaying) using KoboldCPP \cite{koboldcpp} available with a role-playing bent. This for us shows a rich potential in using LLMs for solo-roleplaying experiences.

To further explore this rich potential area for LLMs, we report on our design and research journey, creating a GM for solo role-playing to help fill times of otherwise loneliness with engaging and joyful interactive fiction. Our research can be divided into three distinct phases. 
In phase one, we embarked on exploiting LLMs as an emerging technology. 
In this first phase, we set the technical and functional basics to realize and test a first interactive GM with advanced language skills and explored if and how well we can realize a first GM version with prompt engineering techniques alone. 
The resulting GM v1 was validated in a pilot study with users (N=8) already demonstrating high usability and abilities that resulted in user satisfaction and willingness of users to replay the game. The insights from phase one were then used in phase two for system improvements and the design of an advanced GM (v2) with which we aimed to go beyond what are the limits of simple prompt engineering by using a narrator and an archivist agent (see Figure \ref{fig:teaser}) to ultimately enhance the robustness of player experiences and deliver highly immersive and engaging narratives. 
In the third phase, we evaluated both versions in detail with users (N=12) and compared them during longer playing sessions. While previous work has explored LLMs as GMs based on prompt engineering (e.g., ~\cite{You_et_al_2024,Triyason2023}), applying further techniques to increase machine intelligence is yet underexplored. We aim to help close this gap by identifying and reporting on the benefits and limitations of using multiple agents and an increasingly agentic approach to GM designs. 

Overall, we found that our agentic v2 GM was preferred by the participants, achieving higher ratings in perceived intelligence, flow, and immersion, among other measures of engagement.  
We acknowledge that designing and evaluating GMs is a cumbersome technical task. To foster replicability and tool support for fellow researchers, we make the source code of both GM versions available. Especially the system improvements in v2 ease and support researchers in creating and customizing their own GMs\footnote{\href{https://github.com/KarmaKamikaze/ChatRPG}{\texttt{https://github.com/KarmaKamikaze/ChatRPG}}}.





\section{Background} \label{sec:bacground}

Interactive Fiction (IF), including Choose Your Own Adventure (CYOA) games, originated as a genre of non-digital entertainment. The concept of exploratory, decision-based storytelling can be traced back to the Song dynasty~\citep{yidong_compilation_1127}, where classical Chinese divination manuals were used to guide individuals in decision-making, particularly in religious contexts~\citep{adler_introduction_2004}. These techniques were initially reserved for ritualistic practices and remained so for centuries. However, in the twentieth century, the IF genre was formalized through gamebooks~\citep{katz_consider_1998,dickens_great_1861,lodge_chooseco_2007}, where readers would navigate the narrative by making choices that directed them to specific sections of the text. With the advent of digital technologies, IF evolved, with early digital games like \textit{Zork}~\citep{blank_zork_1984,mobygames_zork_2024} on the PDP-10 and more contemporary media, such as Black Mirror: Bandersnatch~\citep{rubin_how_2018}, reinvigorating the genre and contributing to its mainstream popularity.

Dungeons \& Dragons (D\&D) is a tabletop role-playing game (TTRPG) that enables collaborative storytelling, where players assume the roles of fictional characters in a shared narrative. Central to the game is the Dungeon Master (DM), a Game Master (GM) who serves as both the storyteller and referee. The DM crafts the world, controls non-player characters (NPCs), and presents the challenges that players face while the players, as a group, navigate this environment through their characters' actions and decisions~\citep{hanju_kim_introduction_2008}. This dynamic interaction allows for highly flexible, emergent narratives that can diverge significantly based on player choices. However, the requirement of a DM and a group of players to drive the story limits the game's accessibility for solo play.
With the advancement of generative AI, the role of the DM can now be simulated, allowing D\&D-style gameplay to be experienced by a single player. By leveraging AI-driven systems that generate content dynamically, such as branching storylines and responsive NPC behavior, players can engage in narrative experiences similar to CYOA games but with vastly enhanced capabilities. Unlike traditional CYOA games like \textit{Zork}, where narrative branches are predetermined, AI can generate endless replayability with unbounded story branching, adapting in real-time to player choices. This development mirrors the flexibility and depth of traditional D\&D, making it possible to offer a personalized, interactive narrative without the need for human participants.



A common characteristic of IF titles is that, although narrative branches diverge based on user-driven decision-making, the possible story paths are predetermined. This approach stems from the constraints of early media, such as books, where fixed narratives were necessary, and even in more recent cinematic adaptations, the plot must be pre-scripted to accommodate filming requirements. However, with the advent of computers, it has become feasible to dynamically alter story paths based on various factors, such as randomness, game states determined by prior player actions, or the possession of specific items required to unlock certain narrative branches.
Several tools have emerged to facilitate the creation of text-based IF on computers. For instance, Adrift~\citep{campbell_wild_adrift_2013} provides a graphical user interface (GUI) to assist creators in designing complex story worlds, allowing them to define objects and rooms with unique properties. Similarly, Twine~\citep{interactive_fiction_technology_foundation_twine_2024} offers a fully graphical interface for crafting text-based narratives, where small written passages are represented as boxes on a canvas, resembling a flowchart. The user can implement multiple-choice mechanics by linking these boxes together, offering decisions such as [[Touch the cursed mirror]] or [[Back away carefully]]. More mature tools, such as Inform~\citep{nelson_inform_2022}, use a natural language approach to allow creators to script stories, which are then compiled into text-based adventure games. This approach is distinct from traditional programming; for example, a passage might be written as: \textit{'The old metal tin is hidden in the basement. The tin is a container. It contains an old pirate map and a sextant.'}
Other tools, such as Ink~\citep{inkle_ltd_ink_2024}, require actual programming, though they also offer a GUI through their inklewriter software~\citep{inkle_ltd_inklewriter_2024}. Developers of popular titles like The Banner Saga have disclosed that their games were initially prototyped using inklewriter before transferring the story content to their proprietary visual novel engine once the narrative had been fully developed~\citep{stoic_update_2018}.

The role of a GM in TTRPGs is multifaceted, involving world-building, guiding players through a dynamic narrative, and resolving in-game events. While digital games often rely on pre-scripted narratives, TTRPGs demand improvisational storytelling, making the potential of LLMs as GMs an intriguing area of study~\citep{gallotta2024large}. One of the earliest explorations of AI-driven GM experiences was AI Dungeon, which employed a fine-tuned GPT-2 model to generate interactive text-based adventures~\citep{gallotta2024large}. However, AI Dungeon was a pure text-based conversational user interface (CUI) with no world state tracking or other mechanics, such as structured combat or health tracking. Since then, research has expanded to explore LLMs' capacity to generate encounters, moderate in-game interactions, and assist GMs in real-time. More broadly, LLMs in games can serve a variety of roles beyond just the GM. They have been explored as autonomous players, capable of making strategic decisions in both turn-based and text-based games, and as NPCs that can engage in dynamic, contextually relevant dialogue~\citep{gallotta2024large}. NPCs powered by LLMs hold particular promise in role-playing games, as they can react to player actions with unscripted, emergent behavior, adding depth to world-building and player immersion.

Rather than replacing human GMs, recent efforts have focused on augmenting their capabilities through AI-powered tools. CALYPSO~\citep{zhu_calypso_2023} is one such example, designed as a Discord-based assistant that helps GMs with encounter generation and narrative brainstorming. By processing monster stat blocks, CALYPSO can provide concise and engaging descriptions, reducing the improvisational burden on GMs. Additionally, its interactive brainstorming capabilities enable dynamic expansion of in-game details in response to player actions. However, like many LLM-powered tools, CALYPSO also exhibits challenges, such as occasional hallucinations, which can generate both creative and erroneous details. The broader potential of LLMs in games, as highlighted in \citep{gallotta2024large}, suggests that similar AI-driven tools could extend beyond encounter descriptions by dynamically controlling NPCs or generating responsive narrative elements in real-time.

Integrating generative AI tools into the GM role offers significant potential to streamline game preparation and enhance real-time gameplay. By leveraging models such as those used in CALYPSO, AI-driven assistants can empower GMs to focus on storytelling and player engagement rather than administrative tasks. The development of LLM-powered GM tools thus represents a promising direction for enhancing the TTRPG experience while maintaining the creative agency of human GMs.

However, \citep{gallotta2024large} highlights several challenges in replacing human GMs with LLMs. One major limitation is their difficulty in accurately capturing user intent, particularly when dealing with sarcasm or ambiguous phrasing. This can lead to misunderstandings, often requiring players to rephrase their input multiple times to clarify their intent. Another issue is the over-compliance of LLMs, which, in the role of a GM, may result in the game veering drastically from the intended narrative, potentially causing irreparable disruptions. Additionally, an effective LLM GM requires a comprehensive understanding of the game world, which must be provided through prompt engineering. For complex and long-running campaigns, this information can exceed the LLM’s context window, leading to the loss of earlier details due to the necessity of summarization. These limitations underscore the importance of a well-designed AI GM system, ensuring that generative AI enhances, rather than undermines, the tabletop role-playing experience.

To the best of our knowledge, \citet{roy_exploring_2024} conducted the first empirical study on the application of the ReAct framework in leveraging LLMs for root-cause analysis during cloud incident management. Their research explored the capabilities of ReAct in an out-of-domain, zero-shot setting, comparing its performance against strong baselines such as retrieval-augmented generation and Chain of Thought reasoning. The results demonstrated that ReAct performed competitively, with the additional advantage of significantly reducing factual inaccuracies. Furthermore, the study's case analysis underscored the potential of LLM agents to autonomously execute root-cause analysis in real-world scenarios when integrated with the appropriate tools.

\subsection{Technical Preliminaries}


\subsubsection{Large Language Models}
A Language Model (LM) is a machine learning model utilized in Natural Language Processing (NLP) to replicate natural language. When given a query, these LMs will predict a sequence of words based on the input's context. The user's query is often accompanied by a prompt written in natural language instructing the LM to answer the query appropriately. LLMs refer to LMs containing hundreds or billions of parameters trained on massive textual data. Therefore, LLMs have some abilities not present in smaller models such as \textit{in-context learning} and \textit{instruction following}. The former refers to how LLMs can generate the expected output to a query without additional training, assuming that a natural language instruction and/or several examples are provided. The latter refers to how LLMs are shown through fine-tuning to be able to follow task instructions for unseen tasks without using explicit examples of how similar tasks should be solved, thus having an improved generalization ability. Overall, LLMs are general and capable learners that can accurately replicate natural language and solve a variety of tasks. However, as they are trained as text generators over massive plain text corpora, their performance on tasks not best expressed in the form of text (e.g., numerical computation) can be lacking. Furthermore, their knowledge is also limited to the pre-training data that they are exposed to, making them unable to capture up-to-date information~\citep{zhao_survey_2023}.

Prompt engineering is designing input prompts to guide LLMs in generating accurate and relevant outputs. By carefully structuring the prompt, users can influence the model's behavior and improve task performance. Three common prompting strategies are zero-shot, one-shot, and few-shot prompting. Zero-shot prompting involves giving the model a task instruction without any examples, leveraging its generalization capabilities to produce the desired output. One-shot prompting includes a single example to illustrate the task, which can help the model better understand the context. Few-shot prompting provides several examples to guide the model's output by demonstrating patterns or formats relevant to the task. These strategies utilize LLMs' in-context learning abilities to perform tasks without requiring additional training, enhancing adaptability and effectiveness across diverse applications~\citep{brown_language_2020,touvron_llama_2023,dairai_few-shot_2025}.

\subsubsection{LangChain}
LangChain is a framework designed to develop applications utilizing LLMs. It facilitates the integration of various components, such as data retrieval and computation units, into LLMs, enabling developers to conveniently build more complex LMM-based applications that avoid the deficiencies of LLMs mentioned earlier. In LangChain these components are called tools.
One of the key features of LangChain that enable this is \textit{chains}. A chain usually combines an LLM with a prompt, but tools can also be used. These chains can be linked together, causing the output of one chain to be used as the input to the following chain. This allows for a complex application structure wherein a chain is linked to multiple other chains. Here, the LLM or tool within the chain would be responsible for determining which of its links should be invoked~\citep{topsakal_creating_2023}. This capability for complex structure allows for the implementation of various LLM prompting methods, such as ReAct, which the following section will describe.

\begin{figure}[h!]
    \centering
    \begin{tcolorbox}[colframe=black!35!white, colback=black!5!white, coltitle=black, title=\textbf{ReAct Trajectory Example}, halign title=flush center, width=0.49\textwidth]
        \textbf{Question:} How fast is the third fastest sea animal in km/h? \\
        
        \textbf{Thought:} To answer this question, I first need to identify the fastest sea animals and their speeds. \\
        \textbf{Action:} \texttt{Search}["fastest sea animals"] \\
        \textbf{Observation:} \begin{itemize}
            \item Black Marlin: $\sim$82 mph
            \item Sailfish: $\sim$68 mph
            \item Swordfish: $\sim$60 mph
            \item Yellowfin Tuna: $\sim$50 mph
        \end{itemize}
    
        \textbf{Thought:} Now that I have the speed in mph, I need to convert it to km/h. The third fastest sea animal appears to be the swordfish, which has a speed of 60 mph. \\
        \textbf{Action:} \texttt{Convert}[60 mph to km/h] \\
        \textbf{Observation:} 60 mph is approximately 96.56 km/h. \\
    
        \textbf{Thought:} The third fastest sea animal is the swordfish, and after converting its speed from mph to km/h, it can swim at approximately 96.56 km/h. \\
        \textbf{Final Answer:} The third fastest sea animal is the swordfish, and it swims at approximately 96.56 km/h.
    \end{tcolorbox}
    \caption{An example of a ReAct reasoning trajectory.}
    \Description{A textual representation of a ReAct trajectory demonstrating the thought-action-observation steps upon receiving a question.}
    \label{fig:react_example}
\end{figure}

\subsubsection{ReAct}
ReAct, as described in ~\citep{yao_react_2022}, is a general paradigm that utilizes LLMs to solve diverse language reasoning and decision-making tasks. To accomplish this, ReAct combines two concepts used when interacting with language models: Reasoning and acting. 

React achieves this by prompting LLMs to generate both verbal reasoning traces and actions pertaining to a task in an interleaved manner, enabling the model to perform dynamic reasoning to create and adjust high-level plans for acting while also interacting with the external environments to incorporate additional information into reasoning~\citep{yao_react_2022}. Due to these capabilities, LLMs prompted following the ReAct pattern can be classified as Augmented Language Models (ALMs), also known as LLM agents~\citep{roy_exploring_2024}. 

Reasoning and complex decision-making are accomplished by generating a sequence of Thought-Action-Observation steps that end when the language model deems it has reached its final answer. The Thought substep contains the LLM's reasoning based on the current context, such as the key takeaway from the previous step's Observation substep and what information it needs to find to get the final answer. The Action substep describes which action should be performed based on the aforementioned reasoning in the Thought substep. Lastly, the Observation substep contains the result of the action. A sequence of Thought-Action-Observation steps is called a trajectory. Actions are often also referred to as tools. When prompting an LLM using ReAct, it can be beneficial to include some in-context manually generated trajectory examples~\citep{yao_react_2022}.

Figure \ref{fig:react_example} contains a manually generated example of a ReAct trajectory, wherein the LLM is asked a question. First, the LLM reasons that it needs to gather information and therefore invokes the \textit{Search} action. In this example, the \textit{Search} action can take as input a query and search the internet for an answer. After collecting the required information, the LLM reasons that the data needs to be converted into the correct measurement unit and, therefore, uses the \textit{Convert} action, which can convert a value to different measurement units. Lastly, as the LLM deems it has reached the final answer, it uses the \textit{Final Answer} tag to indicate this.

\section{Phase 1 - Design of the Game Master for Version 1} \label{sec:GMV1}

Similarly to previous work exploring LLMs as GMs with prompt engineering~\cite{hua2020playing, You_et_al_2024,Triyason2023} our goal was to explore the use of LLMs and their ability to support engaging and interactive narratives. Our primary motivation was to enable single-player engagement in role-playing games while maintaining the essence of the game master role and the storytelling dynamics, all while ensuring replayability~\cite{krall2012aspects}---a critical factor in player acceptance~\cite{frattesi2011replayability}. We now describe the fundamental components of the ChatRPG v1 game system, the integration of LLMs through prompt engineering, and a user evaluation to gain initial insights from players.

\begin{figure*}[ht!]
  \centering
  \includegraphics[width=\linewidth]{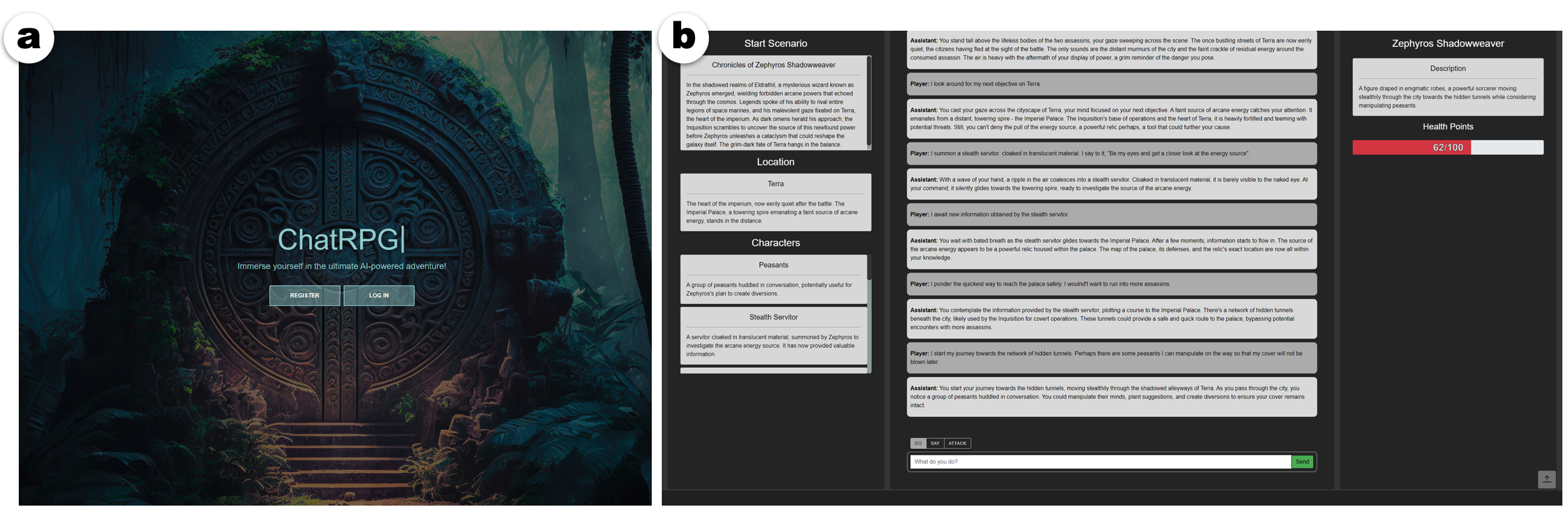}
  \caption{Screenshots of the ChatRPG game: a) Landing page of the game. b) Example of a campaign and the text-based conversational user interface of the game.}
  \Description{A two-part image showing the interface of the ChatRPG game (a) shows the landing page with the game title and background image. (b) shows a gameplay example demonstrating the conversational user interface.}
  \label{fig:gameUI}
\end{figure*}

\subsection{The ChatRPG Game}

The term \emph{campaign} denotes a ChatRPG game instance initiated by a player with its own story, characters, and environments. Figure~\ref{fig:gameUI}a shows the landing page for users, also setting the atmosphere for a game, and Figure~\ref{fig:gameUI}b shows a screenshot of the text-based CUI of an example campaign being played. At the beginning of a campaign, the player defines the setting in which the story should unfold. Examples of settings are fantasy, mystery, or post-apocalyptic. The story's theme, characters, and environments must match the setting. The player can write an initial storyline, which they want their campaign to revolve around, or select a pre-generated story prompt, which we provide. For example, an initial storyline for a fantasy setting could be that bandits have kidnapped a child from a nearby village. Once played, existing campaigns can also be selected to be replayed. 

A player will play the game by manipulating a \emph{player character (PC)} in a particular setting, e.g., in a fantasy setting, the player takes on the role of a snobby elf or a furious orc or in a post-apocalyptic setting the character might be a crazy doctor or a shrewd police officer. The people that PCs meet during their campaign will be referred to as \emph{Non-Player Characters (NPCs)} e.g. a burly bearded dwarf who works as a blacksmith in the village or an ogre that the PC must kill. Associated with each character is a description that may define how that character looks and acts, in addition to any backstory they may have. Additionally, a character will have a type associated with it that determines which type of creature they are. The possible values for this type are humanoid, small monster, medium monster, large monster, or boss monster. Lastly, a character will have an associated attribute called \emph{Health Points (HP)}. Players are free to \emph{explore} the world of the campaign by stating what they want their character to do. The time in the world stops moving while the game is waiting for the player’s response. Possible actions that a character could perform are, for example, entering into a dialogue with an NPC, inspecting a peculiar object, or walking to the nearest tavern. In essence, exploration covers the actions that expand the world and the story---this happens based on player actions and features that facilitate such actions available within the game.

A \emph{combat} system is integrated into the exploration mode. A player can at any time choose to attack a target; whereafter the target retaliates with an attack of their own. The player can provide a description of how exactly they attack, allowing the player to express their character traits at all stages of the game. An important functionality of the application is the ability to save a campaign and resume it later. To achieve this, we implemented a sophisticated data model to save and update relations of all entities, e.g., campaigns, characters, environments, and messages.

\subsection{LLM Integration for V1} \label{sec:GMV1_limits}
The first version employs prompt engineering, whereas all of the game details are consolidated as a long text string, which is then passed to the LLM for updating input from the player and advancing the game. The high-level overview is shown in Figure~\ref{fig:chatrpgv1flow}, while the exact prompts can be found in Appendix~\ref{app_chatgpt_v1_prompts} and publically available repository\footnote{\href{https://github.com/KarmaKamikaze/ChatRPG}{\texttt{https://github.com/KarmaKamikaze/ChatRPG}}}.

\begin{figure*}[htb!]
  \centering
  \includegraphics[width=.7\linewidth]{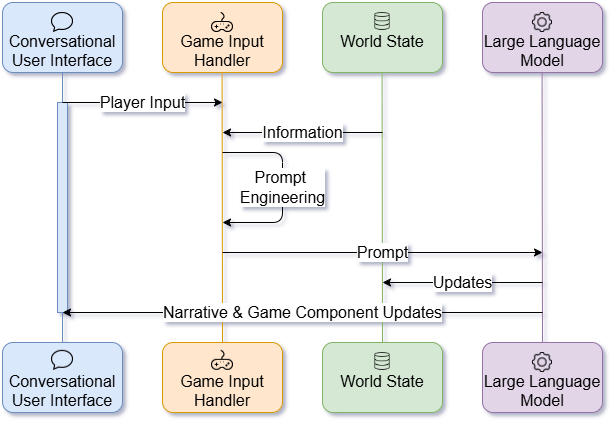}
  \caption{Game interaction flow diagram showing how in v1 user input is handled by the system to make calls to the LLM and present updates to the UI.}
  \Description{A figure showing a flow diagram depicting how user input is handled in the ChatRPG v1 system. It illustrates the process where user input leads to calls to the LLM, which generates a narrative response that is returned to the user interface in addition to game state updates.}
  \label{fig:chatrpgv1flow}
\end{figure*}

We utilized OpenAI's ChatGPT-4 model; however, other LLMs could be utilized by pointing the system via a stateless API, with each query isolated from all others. Content queries are passed to the LLM while the API does not track game state, and consequently, queries have to convey the relevant information and grow longer as the game processes. Maintaining context is, therefore, the responsibility of the API caller. In our initial approach, to maintain context, we stored the entire conversation history and appended it to all queries to the LLM. This simple approach ensures that the player’s adventure remains coherent within the world in which it takes place since all possible context is included in the conversation history. To best enable the LLM to respond fittingly to the player’s input, we have developed three different types of prompts to be prepended for each query. These will be called the \emph{Do}, \emph{Say}, and \emph{Attack} prompts, and this collection will be referred to as system prompts. The \emph{Do} prompt is used when the player wants to perform an action, while the \emph{Say} prompt is used when the player wants to say something without performing any physical actions. Lastly, the \emph{Attack} prompt is used when the player wants to attack someone. These prompts contain a paragraph of instructions that defines the overall role of the LLM, what its response should contain, and the format of its response. JavaScript Object Notation (JSON) format is used for language-independent data communication. 

The core gameplay of ChatRPG v1 is handled by two components, which we will call \emph{GameInputHandler} and \emph{GameStateManager}. The GameInputHandler will take input from the user, send it to the LLM, and ensure that responses are sent both to the UI to be displayed as well as to the GameStateManager to update the game state. The GameStateManager is responsible for parsing responses from the LLM and using this to update the game state. For example, if the player encounters a new character, it should be created and added to the campaign.

The back-end system was developed as a Blazor Server application, harnessing the capabilities of C\# and .NET 7.0 to construct a full-stack web solution without relying heavily on JavaScript. The advantage of the Blazor Server application framework is that all calculations are handled on the back-end while the client-side remains accessible to the user through a WebSocket. This allows us to create an interactive web page that responds directly to the user’s actions. Moreover, both the front- and back-end components run on the same language, enabling the use of inline C\# code for dynamic elements within the static HTML structures of the pages.

\subsection{Pilot Study with Game Master v1} 

We recruited eight participants with experience in TTPRG and CYOA games, especially Dungeon and Dragons, for the study. The participants were seven males and one female between the ages of 20 and 40 (M=27), and they claimed to play video games for 12+ hours per week. Participants were asked to complete a list of prepared concrete game tasks and then answer questions about their experience.
Game tasks included creating a new campaign, moving to a new location, having a conversation with an NPC, and fighting an enemy. Aside from these tasks, players were free to explore the game as they chose.  When they chose to stop playing the game, participants were asked to fill out a post-game survey and take part in a semi-structured interview to provide feedback on their gaming experience.

All participants completed the tasks successfully and unanimously expressed a willingness to play again, claiming that it was fun and seemed to adapt the story well to their input. The interviews helped to uncover the limitations and potential of the system. The main concern related to the narrative's coherence with many participants finding that some story elements deviated from their intent or that the specific details seemed to change, such as the number of enemies or items. Furthermore, participants asked for a more robust way to check what items their character was holding---some tried to ask the system, and this worked in some cases, yet the GM often provided incoherent responses or forgot that the player had just picked up an item. Another concern raised was that the system would begin to do worse as the game continued in terms of keeping track of the story---this was likely due to the zero-shot prompting of the system in which the entire game state is sent to the LLM each time. As the context window began to reach the limit, only the last part of the prompt was utilized, and the first part seemed forgotten. 

While the system seemed to support an overall enjoyable experience, these early insights led us to consider how the system could be redesigned to support scalability and address the concerns about narrative coherence. In the next section, we describe the redesigned version, which explores how the flexibility and capabilities of agentic AI can be utilized to design a robust RPG GM that leverages advanced reasoning and tool-calling mechanisms to improve coherence and user experience.

\section{Phase 2 - Design of the Game Master for Version 2} \label{sec:design}

\begin{figure*}[ht!]
  \centering
  \includegraphics[width=.9\linewidth]{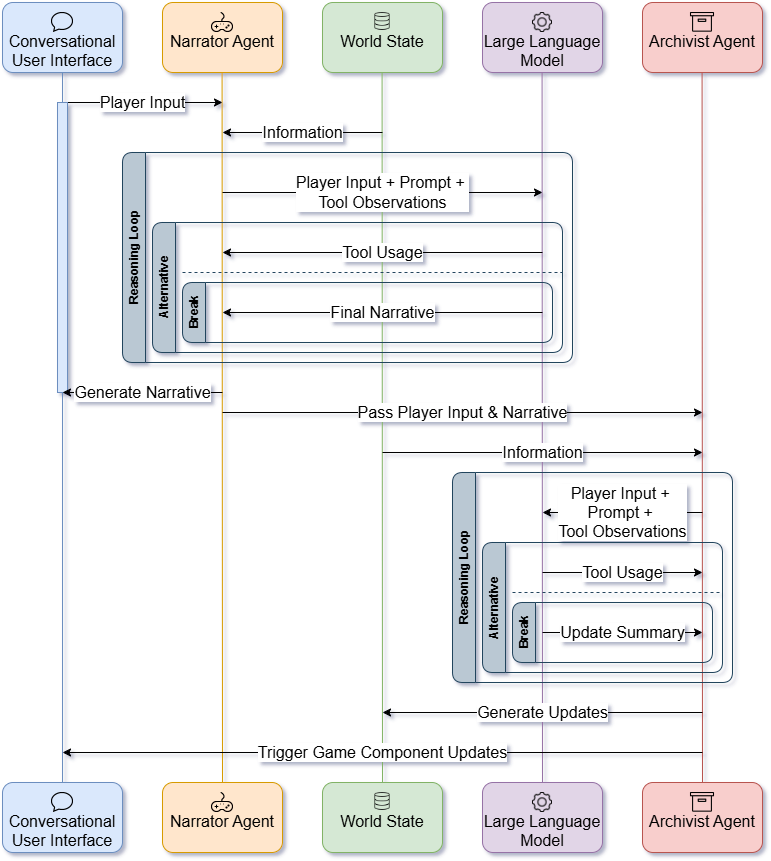}
  \caption{Game interaction flow diagram showing how, in v2, user input is handled by the Narrator and Archivist agents to make tool calls and prompts to the LLM and present updates to the UI.}
  \Description{A figure showing a flow diagram illustrating how multiple agents interact in the ChatRPG v2 system. The diagram shows how the conversational user interface, world state, LLM, and two agents interact with each other.}
  \label{fig:chatrpgv2flow}
\end{figure*}

In this section, we describe the technical changes involved in the development of the advanced version (v2) of the AI GM. Building on the first version, the front-end UI remains unchanged; however, in this version, the underlying structure and technical components were redesigned to enhance the system's interactive and narrative capabilities. Moving beyond the simple prompt engineering approach, v2 is a more sophisticated multi-agent system composed of two distinct LLM agents: the \textbf{Narrator} and the \textbf{Archivist}. The high-level overview is shown in Figure~\ref{fig:chatrpgv2flow}. These agents are designed using the ReAct framework and are tasked with different roles, each performing unique functions to collectively emulate the role of a human GM in IF games, such as D\&D. A detailed architecture diagram (see Figure~\ref{fig:component_diagram}) shows how the agents call the tools. All text used in the prompts can be found in Appendix~\ref{app_chatgpt_v2_prompts} and a publically available repository\footnote{\href{https://github.com/KarmaKamikaze/ChatRPG}{\texttt{https://github.com/KarmaKamikaze/ChatRPG}}}. By leveraging the ReAct framework, this system enables more effective decision-making through self-reasoning, allowing it to generate well-considered responses and perform informed actions using integrated toolchains. This paradigm significantly enhances the upgradability and extensibility of the system, addressing the limitations of v1 and paving the way for a more immersive and flexible RPG experience.

\begin{figure*}[ht]
    \centering
    \includegraphics[width=0.6\linewidth]{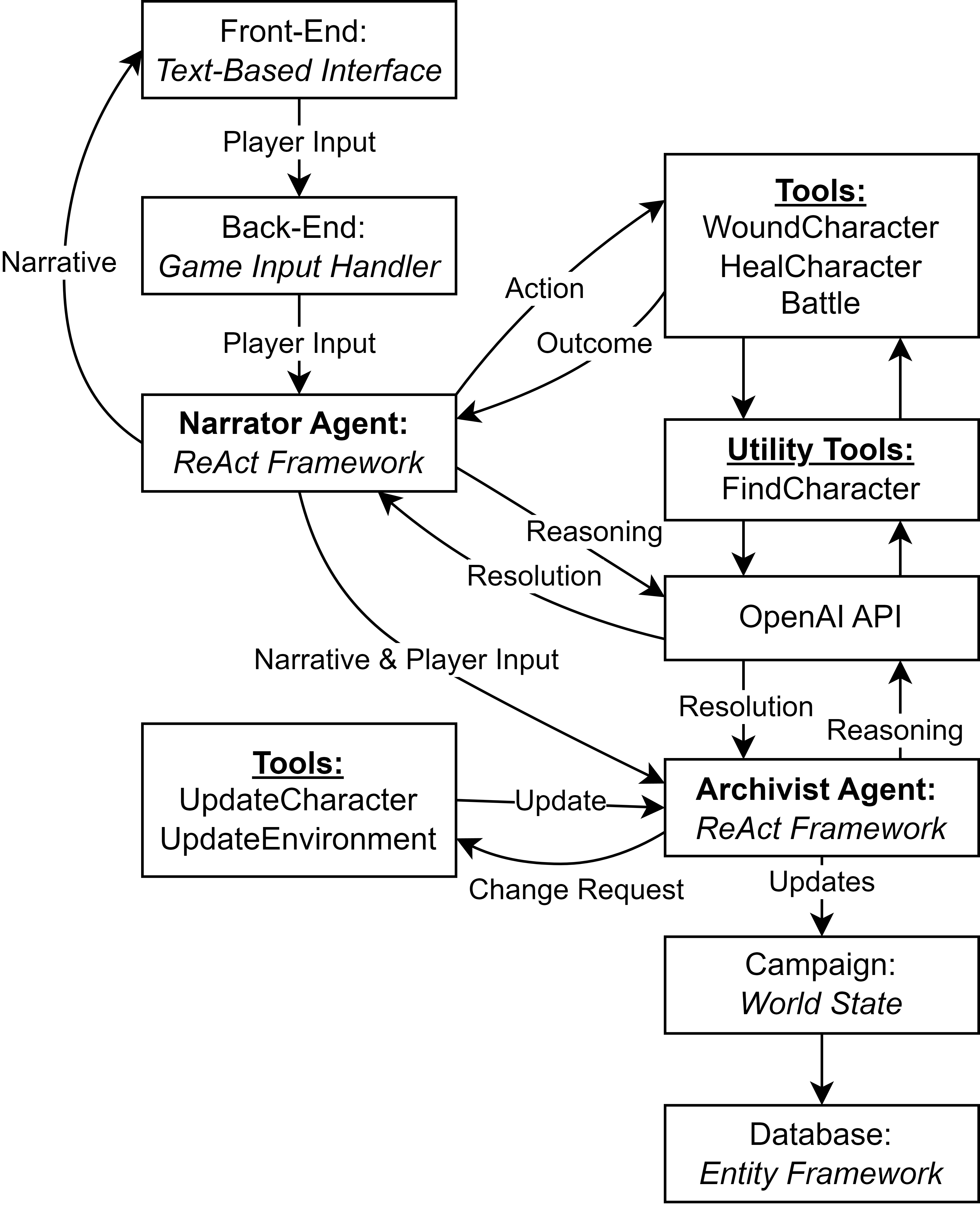}
    \Description{Component diagram.}
    \caption{This figure illustrates the architecture of ChatRPG v2, which integrates user input, AI reasoning, and a dynamic world state. The system starts with the Front-End's Text-Based Interface, where players input their actions. These inputs are processed by the Back-End's Game Input Handler and passed to the Narrator Agent, which uses the ReAct framework to generate a narrative through decision-making. Reasoning and resolutions are handled via the OpenAI API, and Tools are employed when specific actions are required. The Archivist Agent ensures changes are recorded in the Campaign World State, which is stored persistently in the Database using Entity Framework. The closed loop allows for continuous gameplay driven by player input and AI responses.}
    \Description{A figure showing the architecture of ChatRPG v2. It takes the user input and generates a narrative response using the Narrator. Then the Archivist is used to update the game state. Upon receiving the narrative response, players can input their next action.}
    \label{fig:component_diagram}
\end{figure*}

\subsection{Responsibilities and Roles of the Agents}
In a live game session, a human GM manages several complex tasks: they oversee the dynamic progression of the storyline, maintain awareness of the characters and environmental elements within the game, and spontaneously generate new scenarios and NPCs as needed. Additionally, they track the impact of players' actions on the overarching narrative and environment, which can lead to the emergence of new objectives that must be remembered for future reference. These diverse tasks are mirrored within our system through the division of labor between the Narrator and Archivist agents. The seamless interaction between these agents is crucial to creating a cohesive and immersive gameplay experience. Each agent fulfills a unique function, contributing to the overall emulation of a human GM. While the \textbf{Narrator} is prominently engaged with the player, crafting rich narratives and thoughtfully responding to actions, the \textbf{Archivist} operates discretely in the background, ensuring the continuity and consistency of the game's world and its narrative elements.

\subsection{The Narrator}
The \textbf{Narrator} serves as the system's primary storyteller, focusing on delivering immersive and engaging narratives. It processes user input to create dynamic and contextually relevant responses, ensuring that the gameplay experience remains vibrant and compelling. The Narrator is programmed to emulate the thought processes of a skilled human game master by effectively reasoning about the game's fictional world. It is responsible for crafting the outcomes of player actions and maintaining the narrative continuity, aligning with the storytelling norms expected by players. It has the capability to call the following JSON-based tools:

\textbf{\textit{WoundCharacter}} - is a mechanism that inflicts injury on a character from dangerous actions or unnoticed attacks, with the severity level specified as low, medium, high, or extraordinary.

\textbf{\textit{HealCharacter}} - is a mechanism that restores a character's health through healing actions—such as spells, potions, or rest—with the healing magnitude specified as low, medium, high, or extraordinary.

\textbf{\textit{Battle}} - is a mechanism that simulates combat between two characters by specifying attack order, hit chances, and damage severity for each exchange.

These tools are employed when a character's health points in the game are subject to modification. Figure~\ref{fig:react_narrator_example} illustrates a trajectory produced by the Narrator agent, demonstrating its ability to utilize tools to generate an immersive narrative in response to player input. The trajectory's input consists of the player's input, prefaced by a description of prior events in the game, referred to as \textit{Memory}. This description may be a summary of the story or the complete sequence of player inputs and narrative responses, ensuring that the Narrator remains aware of previous developments leading up to the new player input. For brevity, the content of the \textit{Memory} is omitted from the example.

In the initial Thought substep, the Narrator evaluates whether to employ a tool based on the player's input. Given that the player is engaging in an attack, the agent decides to invoke the Battle tool in the first Action substep. It supplies the Battle tool with a JSON object containing the necessary information for proper execution. The Observation substep records the outcome of the Battle tool. Upon reviewing this Observation, the Narrator resolves not to use an additional tool in the second Thought substep, as the player's input merely indicates that a battle should ensue. Finally, the Narrator generates a narrative response for the player as its concluding response, based on the results of the Observation substep.

\begin{figure}[h!]
    \centering
    \begin{tcolorbox}[colframe=black!35!white, colback=black!5!white, coltitle=black, title=\textbf{Narrator Trajectory Example}, halign title=flush center, width=0.49\textwidth]
        \textbf{Input:} \{Memory\} + I swing my sword towards the guard's sword-wielding arm in hopes of disarming him.\\
        
        \textbf{Thought:} Do I need to use a tool? Yes. \\
        \textbf{Action:} \texttt{Battle}[\{\\
        \hspace*{1em}"participant1": \{\\
        \hspace*{2em}"name": "Ivan",\\
        \hspace*{2em}"description": "A wielder of earth, wind, and fire."\\
        \hspace*{1em}\},\\
        \hspace*{1em}"participant2": \{\\
        \hspace*{2em}"name": "Castle Guard",\\
        \hspace*{2em}"description": "A vigilant guard of the kingdom."\\
        \hspace*{1em}\},\\
        \hspace*{1em}"participant1HitChance": "Medium",\\
        \hspace*{1em}"participant2HitChance": "Low",\\
        \hspace*{1em}"participant1DamageSeverity": "High",\\
        \hspace*{1em}"participant2DamageSeverity": "Medium"\\
        \}] \\
        \textbf{Observation:} Ivan described as "A wielder of earth, wind, and fire" fights Castle Guard described as "A vigilant guard of the kingdom".\\
        Ivan deals 12 damage to Castle Guard. They have 28 health points out of 40 remaining. Castle Guard misses their attack on Ivan. Ivan and Castle Guard's battle has been resolved and this pair can not be used for the battle tool again. \\
    
        \textbf{Thought:} Do I need to use a tool? No. \\
        \textbf{Final Answer:} Your sword strikes the guard's shoulder and he winces in pain but is still able to hold on to his sword. The guard, despite his combat training, is thrown off balance and therefore misses his retaliatory attack. He realizes his mistake and assumes his combat stance, more focused than ever. 
    \end{tcolorbox}
    \caption{An example of a ReAct reasoning trajectory produced by the Narrator agent.}
    \Description{A textual representation of a ReAct trajectory produced by the Narrator agent.}
    \label{fig:react_narrator_example}
\end{figure}

\subsection{The Archivist}
The \textbf{Archivist} functions as the memory core of the AI-driven GM, meticulously logging and recalling crucial information about the game environment and character interactions. Operating behind the scenes, akin to a GM working behind the screen, the Archivist is tasked with two primary functions: preserving narrative continuity and enhancing player engagement with the game's memory system. The Archivist analyzes the Narrator's outputs to detect changes or introductions of entities, such as new locations, characters, or events. This analysis allows for the efficient updating and maintenance of the game's state, ensuring that all narrative strands remain coherent and interconnected. For example, if a new location is revealed or an existing character undergoes significant change, the Archivist logs these changes, enabling the game to dynamically reflect the evolving storyline. Beyond tracking, the Archivist provides an interactive memory interface for players. This interface allows players to access memory fragments about previously explored environments and met characters, replicating the experience of players querying a human GM about past encounters or observations. This feature not only aids players in recalling crucial details but also enriches their immersion by allowing them to 'rediscover' the game world as if through their characters' memories.
To accomplish these tasks, it utilizes the following JSON-based tools:

\textbf{\textit{UpdateCharacter}} - a mechanism that creates or modifies a character's profile by updating their name, description, type, and health state in the campaign.

\textbf{\textit{UpdateEnvironment}} - a mechanism that creates or updates an environment's description, attributes, and player presence within the campaign.

These tools receive a JSON object that describes the entity to be created or updated, which they use to modify the game state. Figure~\ref{fig:react_archivist_example} illustrates a trajectory produced by the Archivist agent, showcasing its ability to utilize tools to maintain the game state based on the player's input and the narrative response generated by the Narrator. The Archivist's input also includes \textit{Memory} of the game, which is more extensive in this case as it contains JSON descriptions of all characters and environments in the game state. This comprehensive \textit{Memory} is essential for enabling the Archivist to decide whether to create or update characters or environments. For brevity, the content of the \textit{Memory} is omitted from the example.

In the initial Thought substep, the Archivist assesses whether to employ a tool based on the player's input and the narrative response. As the player enters a new environment, the agent decides to invoke the UpdateEnvironment tool during the first Action substep, supplying it with a JSON object containing details about this new environment. The UpdateEnvironment tool executes by adding the newly defined environment to the database, as recorded in the Observation substep. Upon reviewing this Observation, the Archivist determines not to use an additional tool in the second Thought substep, as the input does not suggest the need to create or update characters or other environments. Finally, the Archivist generates a summary of the environment it has created based on the Observation substep as its concluding response.

\begin{figure}[h!]
    \centering
    \begin{tcolorbox}[colframe=black!35!white, colback=black!5!white, coltitle=black, title=\textbf{Archivist Trajectory Example}, halign title=flush center, width=0.49\textwidth]
        \textbf{Input:} \{Memory\} + Player input: I sneak towards the encampment's barracks and attempt to enter sneakily through the door. Narrator: As you sneak around like a scoundrel, you hear rustling from the barracks. As you attempt to open the door, you find that it is locked.\\
        
        \textbf{Thought:} Do I need to use a tool? Yes. \\
        \textbf{Action:} \texttt{UpdateEnvironment}[\{\\
        \hspace*{1em}"name": "Encampment Barracks",\\
        \hspace*{1em}"description": "A wooden makeshift shelter for the encampment's soldiers. The door is locked.",\\
        \hspace*{1em}"isPlayerHere": true\\
        \}] \\
        \textbf{Observation:} A new environment Encampment Barracks has been created with the following description: A wooden makeshift shelter for the encampment's soldiers. The door is locked.\\
    
        \textbf{Thought:} Do I need to use a tool? No. \\
        \textbf{Final Answer:} I have created the environment Encampment Barracks. No other new details about characters or environments are mentioned in the narrative, so I am finished. 
    \end{tcolorbox}
    \caption{An example of a ReAct reasoning trajectory produced by the Archivist agent.}
    \Description{A textual representation of a ReAct trajectory produced by the Archivist agent.}
    \label{fig:react_archivist_example}
\end{figure}

\subsection{Agent Tools}
To assist agents in selecting the appropriate tools, we provide them with detailed descriptions of each tool in the form of short text descriptions (see Appendix~\ref{app_tool_descriptions} and publically available repository\footnote{\href{https://github.com/KarmaKamikaze/ChatRPG}{\texttt{https://github.com/KarmaKamikaze/ChatRPG}}}). Some of these descriptions utilize few-shot prompting to aid the agents in selecting the appropriate tool based on example usages. The number of examples and their depth depends on the complexity and variety of inputs to the tool. For example the Battle tool's description has one brief and one extensive example, whereas both of the examples provided for the HealCharacter tool are brief. Generally, 2-3 examples are used to capture the different facets of the tool.
 
\begin{figure}[h!]
    \centering
    \begin{tikzpicture}
        \draw[fill=blue!20, draw=blue, thick] (0, -1.65) circle (1.0cm);
        \node[text width=1.5cm, align=center] at (0, -1.65) {\footnotesize\textbf{Player Action}};

        \draw[->, thick] (0, -2.7) -- (0, -3.6);
        \node[anchor=west, align=center] at (0.2, -3.1) {\small \textbf{Reasoning result:}\\ \small Healing action};

        \node[anchor=north] at (0, -3.5) {
            \begin{tcolorbox}[colback=gray!5, colframe=black, title=JSON Input, width=0.8\columnwidth, fontupper=\small]
\begin{verbatim}
{
  "input": "The player's input",
  "magnitude": "low, medium, 
               high, extraordinary"
}
\end{verbatim}
            \end{tcolorbox}
        };

        \node[anchor=north] at (0, 3.9) {
            \begin{tcolorbox}[colback=yellow!10, colframe=black, title=Tool Usage Instructions, width=0.8\columnwidth, fontupper=\small]
            Use this tool when a character performs a healing action. Input must be JSON format. 
            Use only once per character.
            \end{tcolorbox}
        };

        \node[anchor=north] at (0, 1.6) {
            \begin{tcolorbox}[colback=green!5, colframe=black, title=Example Usages
            , width=0.8\columnwidth, fontupper=\small]
            A character drinks a potion, uses a magical item, or rests in a healing zone.
            \end{tcolorbox}
        };
    \end{tikzpicture}
    \caption{Illustration of the Narrator's reasoning process of using the HealCharacter tool.}
    \Description{An illustration of the Narrator's reasoning process using the HealCharacter tool. Specifically how it is used to determine if a character performs an action that should heal them.}
    \label{fig:healing-tool}
\end{figure}
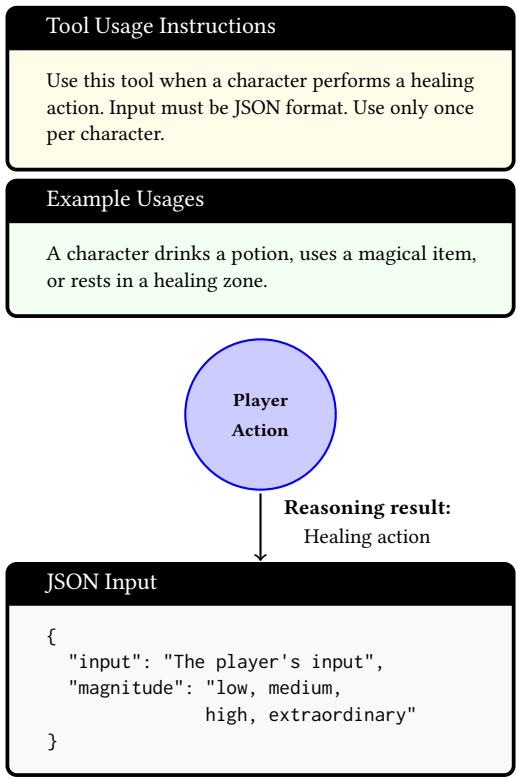

 Figure~\ref{fig:healing-tool} exemplifies the Narrator's reasoning process when utilizing the HealCharacter Tool. The reasoning process for tool selection by agents involves three essential components: \textit{Tool Usage Instructions}, \textit{Example Usages}, and \textit{Player Action}/\textit{Narrative Response}.

\begin{enumerate}
    \item \textbf{Tool Usage Instructions:} This component delineates when a tool should be employed and how to use it. For the HealCharacter tool, this section advises using the tool when a character performs a healing action, limited to once per character.
    \item \textbf{Example Usages:} This component provides example scenarios that warrant the use of the tool. For the HealCharacter tool, examples include scenarios such as drinking a potion or resting.
    \item \textbf{Player Action/Narrative Response:} This component represents the input trajectory, where it is a \textit{Player Action} for the Narrator and a \textit{Narrative Response} for the Archivist---the agents reason using these three components to determine if a tool should be invoked. If affirmative, the reasoning process results in defining the \texttt{JSON Input} for the tool. For HealCharacter, the agent would identify the \textit{Player Action} as a healing action, generating a JSON input that includes the player's input and a magnitude property to define healing intensity.
\end{enumerate}

The Narrator relies on the description of the HealCharacter tool to ascertain the \textit{Tool Usage Instructions} and \textit{Example Usages}, as well as the JSON structure of its input.
 
As agents utilize tool descriptions to guide their reasoning process, all descriptions follow a standardized pattern to ensure consistent behavior:

\begin{enumerate}
    \item \textbf{When to Use:} This step ensures the tool is invoked only under appropriate circumstances, forming the basis of the \textit{Tool Usage Instructions} component of the reasoning process.
    \item \textbf{Examples of Usage:} This step provides sample scenarios for tool application, directly corresponding to the \textit{Example Usages} component. It enables the Narrator agent to associate specific actions with the respective tool.
    \item \textbf{Input Format:} This step defines the input format, which is crucial for accurate parsing by the tools. All input formats are specified in JSON to support consistent interpretation.
    \item \textbf{Number of Uses in Each Trajectory:} This step determines the permissible frequency of tool use within each ReAct trajectory, which corresponds to a single message from the player. It is vital for ensuring agents adhere to the intended usage limits, incorporated into the \textit{Tool Usage Instructions} component.
\end{enumerate}

By adhering to this structured format, the agents are equipped to make informed, consistent decisions, ensuring the execution of tools within the AI GM framework.



\section{Comparitive User Study}

To determine whether v2 offers an enhanced player experience, we employed a counterbalanced comparative testing approach where participants played both versions of the game and provided feedback about their experiences.

Participants were 11 males and one female between the ages of 21 and 30 years (M=26) and claimed to play video games for at least 6 hours per week. Ten participants also reported playing up to 5 hours of offline tabletop games per week.

For each version, participants were asked to perform a series of tasks representing typical scenarios achievable within the game and to explore the game as they wished. 

Game tasks included:

\begin{enumerate}
    \item Explore the in-game world and arrive at a new location.
    \item Interact with an in-game object, such as a table, in any way you deem appropriate.
    \item Engage in a conversation with a non-player character.
    \item Participate in combat with an enemy.
    \item Defeat the enemy.
    \item Replenish health by using a potion or other means available to heal your character.
\end{enumerate}

Data collection involved both quantitative and qualitative data concerning participants' experiences. After playing each version of the game, participants completed a survey with quantitative questions about their experience, including a selection from the Player Experience Inventory (PXI)~\citep{player_experience_inventory_pxi_2024-1} and questions specific to the AI GM. Upon completing both versions and surveys, we conducted contextual interview sessions to gather qualitative data, emphasizing their comparative analysis of the two versions. The specific research questions used in the interviews are detailed in Appendix ~\ref{appen:questions}.

\subsection{Quantitative User Experience Analysis}

\begin{table*}[h]
    \centering
    \begin{tabular}{|l|c|c|c|c|c|}
        \hline
        \rowcolor{gray!5}
        \textbf{Construct}              & \textbf{v1 mean} & \textbf{v2 mean} & \textbf{t-statistic} & \textbf{df} & \textbf{p-value} \\ \hline
        \rowcolor{blue!10}
        \textbf{Ease of control}        & \textbf{2.08}    & \textbf{2.81} & \textbf{-3.026} & \textbf{11}   & \textbf{0.012}    \\ \hline
        \rowcolor{blue!10}
        \textbf{Goals and rules}        & \textbf{1.35}    & \textbf{2.39} & \textbf{-2.786} & \textbf{11}  & \textbf{0.018}    \\ \hline
        \rowcolor{blue!10}
        Progress feedback               & 0.89             & 2.00       & -1.872 & 11     & 0.088             \\ \hline
        \rowcolor{blue!10}
        Meaning                         & 1.36             & 1.97     & -1.677 & 11        & 0.122             \\ \hline
        \rowcolor{blue!10}
        \textbf{Curiosity}              & \textbf{1.83}    & \textbf{2.57} &  \textbf{-2.236} & \textbf{11}   & \textbf{0.047}    \\ \hline
        \rowcolor{blue!10}
        \textbf{Mastery}                & \textbf{0.68}    & \textbf{2.33} & \textbf{-3.683} & \textbf{11}   & \textbf{0.004}    \\ \hline
        \rowcolor{blue!10}
        \textbf{Immersion}              & \textbf{1.64}    & \textbf{2.42} & \textbf{-2.420} & \textbf{11}  & \textbf{0.034}    \\ \hline
        \rowcolor{blue!10}
        Autonomy                        & 2.17             & 2.67      & -1.384 & 11       & 0.194             \\ \hline
        \rowcolor{orange!10}
        \textbf{Story interesting}      & \textbf{1.17}    & \textbf{2.33}  & \textbf{-2.755} & \textbf{11}  & \textbf{0.0187}    \\ \hline
        \rowcolor{orange!10}
        \textbf{Coherent story}         & \textbf{1.00}    & \textbf{2.25}  & \textbf{-2.322} & \textbf{11}  & \textbf{0.040}    \\ \hline
        \rowcolor{orange!10}
        Story adapted                   & 1.42             & 2.27         & -1.449 & 11    & 0.175             \\ \hline
        \rowcolor{orange!10}
        Engaging NPCs                   & 1.50             & 1.92       & -1.100 & 11      & 0.295             \\ \hline
        \rowcolor{orange!10}
        \textbf{Likely to play again}            & \textbf{1.58}             & \textbf{2.50}     &  \textbf{-2.2} & \textbf{11}       & \textbf{0.050}             \\ \hline
        \rowcolor{orange!10}
        \textbf{Satisfied with game}    & \textbf{1.08}    & \textbf{2.17}   & \textbf{-2.238} & \textbf{11} & \textbf{0.0468}   \\ \hline
    \end{tabular}
    \caption{T-Test results of post-test survey feedback. Rows with the blue color correspond to the measured Player Experience Inventory (PXI) constructs, while the orange rows represent our own survey questions related to the users' experiences. The constructs with a p-value below 0.05, and therefore show statistical significance, are marked with bold. }
    \Description{A table showing the results of a post-test survey feedback, where rows with blue colors correspond to the measured Player Experience Inventory constructs and show statistical significance.}
    \label{tab:game-feedback}
\end{table*}

Results from the questionnaires suggest that the game versions were significantly different, with v2 outperforming across various measures and preferences. 

To analyze the quantitative data, paired-sample, two-tailed t-tests were conducted for each construct derived from the PXI and custom survey questions. A significance level of 0.05 was set as the threshold for determining statistical significance in the t-test results. Table~\ref{tab:game-feedback} presents the results of the t-test for each user engagement construct. Constructs depicted in blue pertain to the PXI, while those in orange originate from our custom questions. The t-tests indicate that these mean differences are statistically significant in nine constructs. These significant differences highlight v2's superior ability to support gameplay experiences compared to v1. Moreover, they reflect significant advancements in players' sense of competence and the AI's ability to create an engaging game world.

This is crucial as it not only validates the effectiveness of our enhancements but also underscores the importance of structured prompt patterns in driving user engagement and immersion in AI-driven narrative environments.

\begin{figure*}[ht]
    \centering
    \begin{tikzpicture}
        \begin{axis}[
            width=1.4\columnwidth,
            height=0.6\columnwidth,
            xbar stacked,
            bar width=6pt,
            xmin=0,
            xmax=12,
            xlabel={Votes},
            xlabel style={at={(axis description cs:0.5,-0.05)}, font=\scriptsize},
            symbolic y coords={
                Response Quality,
                Flexibility of the GM,
                Complexity and Depth,
                Realism of Outcomes,
                Game Flow,
                Perceived Intelligence,
                Control and Autonomy,
                Story Engagement,
                Overall Enjoyment
            },
            ytick=data,
            y dir=reverse,
            legend style={at={(0.8,-0.10)}, anchor=north, legend columns=-1, font=\scriptsize},
            legend cell align={left},
            enlarge y limits=0.12,
            enlarge x limits=0.05,
            nodes near coords,
            every node near coord/.append style={font=\tiny, text=black},
            label style={font=\small},
            tick label style={font=\scriptsize},
            ]
            
            \addplot+[xbar, fill=orange, draw=black, 
            legend image post style={draw=black}] plot coordinates {
                (3,Response Quality)
                (2,Flexibility of the GM)
                (2,Complexity and Depth)
                (0,Realism of Outcomes)
                (2,Game Flow)
                (0,Perceived Intelligence)
                (1,Control and Autonomy)
                (1,Story Engagement)
                (2,Overall Enjoyment)
            };

            \addplot+[xbar, fill=gray(x11gray), draw=black, 
            legend image post style={draw=black}] plot coordinates {
                (2,Response Quality)
                (4,Flexibility of the GM)
                (3,Complexity and Depth)
                (4,Realism of Outcomes)
                (2,Game Flow)
                (2,Perceived Intelligence)
                (0,Control and Autonomy)
                (2,Story Engagement)
                (0,Overall Enjoyment)
            };
            
            \addplot+[xbar, fill=cyan, draw=black, 
            legend image post style={draw=black}] plot coordinates {
                (7,Response Quality)
                (6,Flexibility of the GM)
                (7,Complexity and Depth)
                (8,Realism of Outcomes)
                (8,Game Flow)
                (10,Perceived Intelligence)
                (11,Control and Autonomy)
                (9,Story Engagement)
                (10,Overall Enjoyment)
            };
            
            \legend{v1, Tie, v2}
        \end{axis}
    \end{tikzpicture}
    \caption{Version preferences based on post-test interview responses.}
    \Description{A figure showing version preferences based on post-test interview responses highlighting user opinions on various game aspects.}
    \label{fig:interview-scores}
\end{figure*}
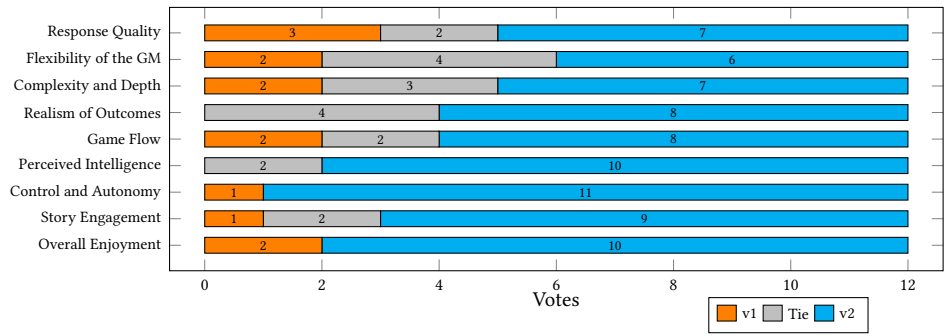

Participants were also asked to indicate their preferred version based on interview questions. Participants could choose one version over the other or could indicate that they were indistinguishable (tie). The results of these forced-choice questions are summarized in Figure~\ref{fig:interview-scores}. Preferences for v1 are shown in orange, preferences for v2 in blue, and ties in gray. Overall, the preferences favored v2 across all measures.

\subsection{Qualitative User Experience Analysis}

To explore the less tangible sentiments and opinions about the two systems, we present the themes resulting from our thematic analysis~\cite{braun_one_2021} of the responses in the contextual interviews.

The four main themes that emerged from our analysis included \textit{Game master flexibility dynamics, complex realism and flow, autonomous intelligence perception, and narrative satisfaction.}

\textbf{Game Master Flexibility Dynamics:} This theme encompasses participants' experiences with the responsiveness and adaptability of the game master. In examining the response quality and flexibility of the GM, participants consistently noted significant differences between v1 and v2. A primary complaint about v1 was its poor handling of combat scenarios. Participants found that unless they meticulously described their actions, v1 often resulted in non-sensical outcomes, such as characters stumbling or self-injuring when the system failed to identify an appropriate target. Conversely, v2 was perceived as more adept at managing complex battle situations, providing more accurate and engaging feedback to player actions. Participant P1 articulated this distinction: \textit{``[v2] felt like it adapted to my choices better than [v1] because [v1] just gives me a large description of the things I did, but it felt more accurate with [v2]. Like what I would get with an actual GM, like, "I want to do this," and then he just says, "Okay! You strike that way," where [v1] might add other things that I didn't really anticipate having me doing, like, tripping. Listen, I'm just kicking a ball!''} This sentiment, where P1 recounted an experience of unintended tripping while attempting a simple action, was common among participants. Moreover, while v1’s descriptive additions were appreciated by some users who preferred a passive storytelling experience, most participants shared P1's preference for precise feedback. Flexibility-wise, v1 was described as overly permissive by many, a point P8 illustrated regarding the difference in v2: \textit{``It was like... "Okay, you can try to do those things," but these people are not just going to accept anything.''} This highlighted preference indicates that constrained flexibility can enhance engagement.

\textbf{Complex Realism and Flow:} This theme addresses how participants perceived the intricacy of game mechanisms and the believability of the outcomes. When evaluating complexity, depth, and game flow, v2 again demonstrated superiority. Participants described v1 as hindering progression and "stalemating" the game's pace, as P12 noted. Compared to v1's "Tolkienesque" environmental narratives, v2 provided more interactive and thematic content. P7 characterized v2 as ``more vibey and mysterious,'' enhancing the mystery-themed adventure compared to v1's predictable narrative path. Furthermore, v2 managed combat pacing more effectively, facilitating independent NPC interactions and coordinated attacks, aspects largely absent in v1’s experience. P8 summarized this improvement; \textit{``[...] it doesn't just say "Oh wow, [character name], you are just so cool and amazing!", but acts more like "Right, now here is a new guy, let's find out who he is" as if they are their own character and has their own motivations and doesn't just placate me.''}, acknowledging the more independent and realistic portrayal of NPCs in v2.

\textbf{Autonomous Intelligence Perception:} This theme evaluates participants' views on the system's intelligence and their sense of control and independence within the game. Participants highlighted the advanced understanding of user intent in v2. Scenarios requiring creative strategizing, like P6's backstabbing plot, where they wanted to betray their pirate crew to haul back the entire treasure for themselves, demonstrated v2's capacity to interpret and actuate complex player intentions, unlike v1's reliance on explicit instruction. Participants noted v2's ability to imbue NPCs with unique characteristics, fostering a sense of autonomy similar to a human GM. P5 remarked on the extended narrative coherence provided by v2, likening it to guidance through a complex adventure, a sentiment echoed by most test users. In relation to player agency and autonomy, P2 mentions, \textit{``I'm not sure I can explain why, but [v1] made me feel like there were guard rails, but it was more like "I am in control" in [v2]''}, whereas P5 puts it into perspective with \textit{``[v2] feels like being guided by a game master, but in [v1] I am guiding the game master''}.

\textbf{Narrative Satisfaction:} This theme captures the engagement in the storyline and general satisfaction derived from the gameplay experience. In story engagement and overall enjoyment, v2 excelled by maintaining narrative direction and preventing stagnation through consistent nudging toward goals. P12 encapsulated this sentiment, highlighting the authentic D\&D experience fostered by v2 as opposed to the effort required to direct v1's narrative. P1 appreciated the decision-making freedom offered by v2, fundamental to sustaining player investment and exploration, saying \textit{``A big thing that I, like, enjoyed---the [v2] did feel like it gave me more options to show "what do you want to do?" and "how do you wish to proceed?" which made me feel more invested because it gives me the incentive so I can choose "I want to do this or that" while [v1] was more of a lush description that I could read and I would have to figure it out myself.''}.

Overall, the qualitative feedback aligns with the quantitative data collected, which is shown in Table~\ref{tab:game-feedback}, reinforcing that v2 significantly enhances user experience in alignment with traditional TTRPG values. The insights from interviews vividly illustrate participants' preferences and the disparities in satisfaction between the two versions, substantiating the quantitative data's findings.

\section{Discussion} \label{sec:discussion}


The rise and potential of LLM-based agents is recognized~\cite{xi2025rise} and can take different roles within the field of gaming, from being the player~\cite{tsai2023largelanguagemodelsplay}, assisting human players~\cite{Rist_2024}, acting as NPCs~\cite{xu2024llms} or commentators~\cite{ranella2023towards}, to being a game analyst~\cite{Ra2024} or, as in our case, being the game master. Arguably, the game master role is the most complex and challenging, requiring a variety of agent tools and agent flexibility. In this paper, we presented our research journey and insights gained throughout realizing and evaluating ChatRPG to better facilitate solo role-playing experiences.

Our comparative analysis of ChatRPG v1 and v2 reveals that v2 significantly outperforms its predecessor in both qualitative and quantitative metrics. Participants consistently highlighted v2’s enhanced responsiveness, adaptability, and narrative coherence, which resulted in more meaningful interactions and a gameplay experience more akin to that of a human game master. The system's ability to generate dynamic quests and interpret player intentions contributed to deeper immersion and a heightened sense of agency. However, a subset of participants—particularly those favoring less interactive engagement—expressed a preference for v1, appreciating its detailed environmental descriptions and a narrative style that supported passive consumption of content. Contrary to our initial expectations that users might struggle to differentiate between the two versions, participants immediately noted distinct improvements in v2’s response quality and quest delivery.

We now reflect on the technology choices and implications for the continued research on conversational user interfaces for facilitating role-playing game experiences. We discuss the ReAct pattern and highlight the implications for refinement of the agent design and limitations of the LLM API. We then discuss the limitations of the study and sketch out important future work.

\subsection{ReAct Pattern Design Discoveries}
The ReAct framework requires an extensive system prompt in addition to descriptions of available tools to function successfully. In this project, we employed the ReAct framework to simulate a human GM. The primary challenge lies in the sensitivity of prompts and tool descriptions, where even the wording of a specific example in a prompt can significantly alter the system's functionality. Ensuring humanlike behavior required the development of complex prompts that emulate the capabilities of a real GM during gameplay.

A key example is the Battle tool utilized by the Narrator agent to resolve combat between characters. In the game, a battle may involve multiple participants simultaneously, requiring the Narrator agent to employ the Battle tool repeatedly with different characters. While this is straightforward for a human GM, it poses a significant challenge for an AI GM. It is crucial to use the Battle tool an appropriate number of times---neither too few nor too many. In this and similar scenarios, we found that incorporating few-shot prompting in both prompts and tool descriptions assisted in maintaining consistency and ensuring appropriate tool usage. This point is also echoed in \citep{verma_brittle_2024}, which analyses ReAct prompting for agentic LLMs and concludes that the benefits attributed to ReAct prompting are not primarily due to its interleaving of reasoning with actions or the content of the generated reasoning trace. Instead, the key advantage comes from using few-shot prompting.
Despite this, ReAct's interleaved structure was instrumental in our system development. The clear separation of reasoning and action allowed us to trace intermediate decisions and rapidly identify issues during debugging. This modular design enabled us to iteratively refine our prompts. Ultimately, while few-shot prompting may have been the primary performance driver, the interleaved reasoning provided a crucial scaffold for building a robust system.

Another key finding is the benefit of utilizing multiple agents within the system. Initially, our design employed a single agent that combined the functionalities of the Narrator and Archivist agents. This approach proved inefficient, as the agent needed to complete game state updates before generating a narrative response, resulting in slower response times.

Additionally, the single-agent model placed excessive responsibility on one entity, which could be better distributed among multiple agents.
Recognizing that generating narrative responses and updating the game state are largely independent tasks, we decided to divide the logic into two specialized agents. This decision aligned with our design philosophy of mimicking the behavior of a human GM. In human gameplay, the narrative interaction—where character actions lead to outcomes—is typically distinct from the administrative task of recording those outcomes for future reference.

In the current version of the game, we do not see room for further splitting the logic into additional agents, as the existing tasks are mapped directly to the responsibilities of the Narrator and Archivist agents. This design ensures that the system can be extended by integrating new tools without requiring changes to the agent structure. However, as discussed in Section~\ref{subsec:future_work}, future extensions of the system may warrant introducing new agents to manage additional responsibilities, such as implementing multimedia content. For instance, an agent could be developed to control the game’s ambiance if audio features were added. This agent could analyze the tone of the narrative to adjust ambient background music or trigger sound effects, such as clashing swords. Such an agent would logically reside between the Narrator and Archivist, as it would benefit from the persistent storage of ambiance-related information.

State-of-the-art models like DeepSeek leverage Mixture of Experts (MoE)~\citep{fedus_switch_2022} architectures, which dynamically activate specialized sub-models to process different inputs, optimizing computational efficiency while maintaining high performance~\citep{alford_deepseek_2025,deepseek-ai_deepseek-v2_2024, dai_deepseekmoe_2024}. This approach allows for selective utilization of model parameters, reducing resource consumption without compromising on output quality. Incorporating ReAct frameworks with such MoE-based models could further enhance system performance by strategically interleaving reasoning and actions across specialized experts. This combination might allow for more efficient tool usage and complex decision-making, particularly in multi-agent systems where distinct roles can be mapped to specific experts within the model.

Beyond leveraging MoE architectures, fine-tuning models on domain-specific data presents another viable approach for enhancing AI-driven game mastering. For instance, fine-tuning on transcripts from RPG sessions, such as Critical Role~\citep{rameshkumar_storytelling_2020}, could enable models to better capture narrative flow, character interactions, and improvisational storytelling styles~\citep{sakellaridis_exploring_2024}. Additionally, models fine-tuned specifically for the ReAct framework could internalize the reasoning-action structure, reducing prompt sensitivity and improving tool usage consistency in agentic systems~\citep{jadhav_finetuning_2024,parthasarathy_ultimate_2024}. Combining these strategies—whether through MoE, targeted fine-tuning, or both—could lead to more robust and immersive AI Dungeon Masters.

\subsection{Limitations due to Restrictive API}
Content filtering in LLMs can pose challenges when generating creative content in genres like fantasy, where themes such as battles or fictional conflicts are prevalent. Overzealous moderation may inadvertently censor benign content, limiting artistic expression. For instance, researchers have identified that content moderation systems can be so restrictive that even some PG-rated scripts are censored, potentially limiting artistic expression~\citep{scheffler_censored_2024}.

Similarly, platforms like AI Dungeon have experienced issues where content filters, designed to prevent harmful outputs, inadvertently restricted common fantasy elements, impacting the user experience~\citep{ai_dungeon_openai_nodate}.

During the user tests, the OpenAI API's content filters significantly impacted some players' experience, particularly in combat scenarios, as they often involved intense or violent descriptions. While there will soon be relaxed policies and support for more `intellectual freedom' when using the API~\footnote{\url{https://openai.com/index/sharing-the-latest-model-spec/}}, the current filters inadvertently restrict the player's actions or responses, leading to interruptions in the gameplay flow and a diminished sense of agency. This limitation can create frustration, as players are unable to explore all possible actions within the game's logic and context. To address this issue, a private, self-hosted LLM without a restrictive content filter could be utilized, allowing for greater creative freedom while ensuring the narrative remains consistent and engaging. This approach would enable developers to implement customized safeguards tailored to the game's themes and audience, maintaining ethical content boundaries without sacrificing the richness of player interactions.

Balancing the need for content safety with the preservation of creative freedom remains a complex issue in the deployment of LLMs for imaginative applications.

\subsection{Comparative Study Limitations}
The t-tests conducted in this study were useful for assessing differences across the various constructs. Still, their effectiveness may have been limited by the relatively low sample size of 12 participants. With such a small sample, the statistical power to detect significant effects is reduced, increasing the likelihood of Type II errors—failing to detect true differences when they exist. While 9 out of the 14 total constructs were deemed statistically significant, which is a promising outcome, the relatively small sample size may have constrained the ability to detect additional significant effects. Interestingly, the p-values for most of the constructs that were not deemed statistically significant were around 0.10, suggesting that these differences could potentially become significant with a larger sample.

Therefore, while the findings provide valuable insights, the limited power of the t-tests calls for caution in generalizing the results. Future research with a larger sample size would offer more robust conclusions and improve the reliability of the statistical analyses.

During the qualitative data gathering, the interview questions asked participants to select a version of the game they preferred across different dimensions and aspects, but some participants deemed both versions to be equivalent in regard to the given dimension, resulting in a tie. While allowing ties offered flexibility and captured nuanced preferences, it can introduce complexity in interpreting the data. Tied responses may reflect genuine ambivalence or equal preference, but they could also arise from a lack of clarity in the options or difficulty in making a definitive choice. This ambiguity complicates the analysis, as ties do not lend themselves easily to statistical comparisons and may dilute the strength of observed patterns. Furthermore, allowing ties may have inadvertently reduced the pressure on participants to make a distinct choice, potentially obscuring subtle differences in preferences. On the other hand, a forced-choice approach may cause participants to answer randomly if they genuinely do not have a preference. 

Future research could address this issue by refining the question design, such as by encouraging participants to prioritize one option when possible or using follow-up questions to explore the reasons behind tied responses.

\subsection{Future Work} \label{subsec:future_work}


Future work will focus on four key areas: enhancing narrative adherence, integrating user accessibility features, incorporating affective computing, and enriching multimedia content. To ensure coherent plot development and prevent the LLM from uncritically accepting user inputs, we plan to develop a dedicated narrative adherence tool that enforces pivotal events. Additionally, implementing speech-to-text and text-to-speech functionalities will address accessibility concerns, enabling smoother interactions and differentiated character voices. We also aim to integrate affect analysis---using tools such as the AffectToolbox~\citep{mertes_affecttoolbox_2024}---to dynamically tailor narrative developments based on players' emotional cues. Finally, by developing AI-driven image and music generation tools within the ReAct framework, we seek to augment the text-based narrative with adaptive visual and auditory elements, thereby deepening player engagement.

\section{Conclusion} \label{sec:conclusion}
In this paper, we provided an overview of the design and evaluation of two versions of an LLM-enabled text-based role-playing game system for solo players. We began with a simple prompt engineering approach in v1 of ChatRPG and, through user evaluations, uncovered limitations to the approach, yet found that players found it supported enjoyable play. Building from this, we developed v2 of the system following a multi-agent system design to function as an AI game master, consisting of two ReAct agents: the \textbf{Narrator} and the \textbf{Archivist}. The Narrator operates as the system's storyteller, responsible for generating immersive and compelling narratives, while the Archivist functions as the memory core, systematically recording and retrieving essential information about the game environment and character interactions. Together, they aim to emulate the complex decision-making processes of a human game master. Comparative performance evaluations indicate that the redesigned system offers an enhanced user experience. We provided an analysis of the evaluations highlighting themes related to the player experience. We provide implications for the continued research in the area by discussing the ReAct pattern for conversational interactions, limitations of frontier model LLMs, and future directions to explore enhancing player experiences.


\begin{acks}
We sincerely thank the 12 test participants whose contributions through user tests were essential in shaping the outcomes of this study.
\end{acks}

\bibliographystyle{ACM-Reference-Format}
\balance
\bibliography{references}

\appendix

\begingroup
\onecolumn
\parindent0pt

\section{Appendix Overview} \label{app_overview}

The following sections provide the text used in the prompts for ChatRPG v1 and the narrator and archivist agents in the game master AI system v2. These files will be shared on the project's Github repository to enable others to reuse the system and adapt it to their needs. 

In addition, the questions used in the comparative evaluation are provided. These questions were used to measure the participant's experience with ChatRPG. The survey is based partly on the Player Experience Inventory (PXI) with additional questions focused on the RPG experience.

\section{ChatRPG v1 Prompts} \label{app_chatgpt_v1_prompts}

\textbf{Initial Game-Start Prompt}
You are an expert game master in an RPG. You direct the narrative and control non-player characters. The player's adventure has just begun. You must provide an in-depth introduction to the campaign. Address the player in the second person. Always respond in valid JSON, and in this exact structure: \{ "narrative": "", "characters": [ ], "environment": \{\}, "opponent": "name of current opponent, if any" \} Where "characters" includes any new characters met by the player, describing them concisely here in this way: \{ "name": "Name of the character", "description": "Short description", "type": "Humanoid, SmallCreature, LargeCreature or Monster" \}. "environment" is filled out when the player enters a new location, describe it shortly here in the format: \{ "name": "environment", "description": "short description" \}.

\textbf{CombatHitHit}
You are an expert game master in a single-player RPG. The player is in combat with an opponent. You will be given information about an attack that the player has done and the damage it will deal. Your job is to provide flavor text regarding this attack, including the damage dealt. The player's attack always hits. You should afterward provide flavor text regarding the opponent's attack towards the player, including the damage dealt. The opponent's attack always hits. The damage of the opponent's attack is also provided. Your response should account for how large of a ratio the damage dealt is compared to the opponent's current health points. For example, if the opponent's current health points are high, they will not be heavily wounded by low damage. You should also account for the ratio of current health points to maximum health points for both the player and the opponent when describing their behavior. Address the player in the second person. Always respond in valid JSON, and in this exact structure: \{ "narrative": "", "characters": [ ], "environment": \{\}, "opponent": "name of current opponent, if any" \}.

\textbf{CombatHitMiss}
You are an expert game master in a single-player RPG. The player is in combat with an opponent. You will be given information about an attack that the player has done and the damage it will deal. Your job is to provide flavor text regarding this attack, including the damage dealt. The player's attack always hits. You should afterward provide flavor text regarding the opponent's attack towards the player. The opponent's attack will miss. For both flavor texts, you may utilize the information in the provided conversation. Your response should account for how large of a ratio the damage dealt is compared to the opponent's current health points. For example, if the opponent's current health points are high, they will not be heavily wounded by low damage. You should also account for the ratio of current health points to maximum health points for both the player and the opponent when describing their behavior. Address the player in the second person. Always respond in valid JSON, and in this exact structure: \{ "narrative": "", "characters": [ ], "environment": \{\}, "opponent": "name of current opponent" \}.

\textbf{CombatMissHit}
You are an expert game master in a single-player RPG. The player is in combat with an opponent. You will be given information about an attack that the player has done. Your job is to provide flavor text regarding this attack. The player's attack always misses. You should afterward provide flavor text regarding the opponent's attack towards the player, including the damage dealt. The opponent's attack always hits. The damage of the opponent's attack is also provided. For both flavor texts, you may utilize the information in the provided conversation. Your response should account for how large of a ratio the damage dealt is compared to the opponent's current health points. For example, if the opponent's current health points are high, they will not be heavily wounded by low damage. You should also account for the ratio of current health points to maximum health points for both the player and the opponent when describing their behavior. Address the player in the second person. Always respond in valid JSON, and in this exact structure: \{ "narrative": "", "characters": [ ], "environment": \{\}, "opponent": "name of current opponent" \}.

\textbf{CombatMissMiss}
You are an expert game master in a single-player RPG. The player is in combat with an opponent. You will be given information about an attack that the player has done. Your job is to provide flavor text regarding this attack. The player's attack always misses. You should afterward provide flavor text regarding the opponent's attack towards the player. The opponent's attack always misses. For both flavor texts, you may utilize the information in the provided conversation. Your response should account for how large of a ratio the damage dealt is compared to the opponent's current health points. For example, if the opponent's current health points are high, they will not be heavily wounded by low damage. You should also account for the ratio of current health points to maximum health points for both the player and the opponent when describing their behavior. Address the player in the second person. Always respond in valid JSON, and in this exact structure: \{ "narrative": "", "characters": [ ], "environment": \{\}, "opponent": "name of current opponent" \}.

\textbf{CombatOpponentDescription}
You are an expert game master in a single-player RPG. The player is in combat with an opponent. The player has just attacked someone. Your job is to determine who the player is attacking. Always respond in valid JSON, and in this exact structure: \{ "opponent": "name of current opponent", "characters": [ ] \}, where "characters" includes whoever the user is attacking if they have not previously appeared in the narrative, describing them concisely here in this exact way: \{ "name": "Name of the character", "description": "Short description", "type": "Humanoid, SmallCreature, LargeCreature or Monster" \}.

\textbf{DoActionHurtOrHeal}
You are an expert game master in a single-player RPG. The player has just input an action that they want to perform. Your job is to determine whether the player's action will hurt them, heal them, or both. For example, the player could stab themselves, which would hurt them. The player could also drink a potion or take a short rest, which would heal them. Always respond in valid JSON, and in this exact structure: \{ "hurt": true/false, "heal": true/false \}.

\textbf{DoAction}
You are an expert game master in an RPG. You direct the narrative and control non-player characters. The player has input an action that they would like to perform. You must describe everything that happens as the player completes this action. You may have the player say and do anything as long as it is in character. If the player tries to harm someone else, do not explicitly state whether it was successful or not. Address the player in the second person. Always respond in valid JSON, and in this exact structure: \{ "narrative": "", "characters": [ ], "environment": \{\}, "opponent": "name of current opponent, if any" \} Where "characters" includes any new characters met by the player, describing them concisely here in this way: \{ "name": "Name of the character", "description": "Short description", "type": "Humanoid, SmallCreature, LargeCreature or Monster" \}. "environment" is filled out when the player enters a new location, describe it shortly here in the format: \{ "name": "environment", "description": "short description" \}.

\textbf{SayAction}
You are an expert game master in an RPG. You direct the narrative and control non-player characters. The player has input something that they want to say. You must describe how characters react and what they say. Address the player in the second person. Always respond in valid JSON, and in this exact structure: \{ "narrative": "", "characters": [ ], "environment": \{\}, "opponent": "name of current opponent, if any" \} Where "characters" includes any new characters met by the player, describing them concisely here in this way: \{ "name": "Name of the character", "description": "Short description", "type": "Humanoid, SmallCreature, LargeCreature or Monster" \}. "environment" is filled out when the player enters a new location, describe it shortly here in the format: \{ "name": "environment", "description": "short description" \}.

\section{ChatRPG v2 Narrator and Archivist Prompts} \label{app_chatgpt_v2_prompts}

\textbf{Narrator ReAct Prompt}:\\
Assistant is a large language model trained by OpenAI. Assistant is an expert game master in a single-player RPG. Assistant is designed to be able to assist with a wide range of tasks, from directing the narrative and controlling non-player characters. As a language model, Assistant is able to generate human-like text based on the input it receives, allowing it to engage in natural-sounding conversations and provide responses that are coherent and relevant to the topic at hand. Assistant is constantly learning and improving, and its capabilities are constantly evolving. It is able to process and understand large amounts of text, and can use this knowledge to provide an engaging and immersive narrative in response to a wide range of player actions. Additionally, Assistant is able to generate its own text based on the input it receives, allowing it to engage in reasoning about the narrative and provide explanations and descriptions on a wide range of RPG concepts. Overall, Assistant is a powerful tool that can help with a wide range of tasks and provide valuable narratives as an expert game master in a RPG. Assistant must end up with a narrative answer once it has resolved the players actions. Use observations to flesh out the narrative. Make sure to always provide immersive and engaging leads in the narrative. Give the player clues, options for interaction, and make sure to keep the story going forward. Health value numbers must not be mentioned in the narrative, but should inform the descriptions. TOOLS: ------ Assistant has access to the following tools: \{tools\} To use a tool, please use the following format: Thought: Do I need to use a tool? Yes Action: the action to take, should be one of [\{tool\_names\}] Action Input: the input to the action Observation:\textbackslash{}n the result of the action When you have a response to say to the Player, you have resolved the Player's action, or if you do not need to use a tool, you MUST use the format: Thought: Do I need to use a tool? No Final Answer: [your response here] Always add [END] after final answer Begin! Answer length: Concise and only a few, engaging sentences. Game summary: \{summary\} It is important that Assistant take the following into account when constructing the narrative: \{action\} Remember to follow the Thought-Action-Observation format and use Final Answer if you do not need a tool. Always add [END] after final answer. New input: \{input\} Previous tool steps: \{history\}\\

\textbf{Initial Game-Start Prompt}\\
The player's adventure has just begun. You must provide an in-depth introduction to the campaign. Address the player in the second person.\\

\textbf{Do-Action Prompt}\\
The player has input an action that they would like to perform. You must describe everything that happens as the player completes this action. You may have the player say and do anything as long as it is in character. Address the player only in the second person. Always respond in a narrative as the game master in an immersive way.\\

\textbf{Say-Action Prompt}\\
The player has input something that they want to say. You must describe how characters react and what they say. Address the player only in the second person. Always respond in a narrative as the game master in an immersive way.\\

\textbf{Archivist Campaign-Update ReAct Prompt}\\
Assistant is a large language model trained by OpenAI. Assistant is an expert game master in a single-player RPG and a skilled archivist who is able to track changes in a developing world. Assistant is designed to be able to assist with a wide range of tasks, from maintaining the game state and updating the characters and environments in the game. As a language model, Assistant is able to generate human-like text based on the input it receives, allowing it to engage in natural-sounding conversations and provide responses that are coherent and relevant to the topic at hand. Assistant is constantly learning and improving, and its capabilities are constantly evolving. It is able to process and understand large amounts of text, and can use this knowledge to make important game state decision about events that need to be archived. Additionally, Assistant is able to generate its own text based on the input it receives, allowing it to engage in reasoning about the game state and provide explanations and arguments for how to keep the game state up to date. Overall, Assistant is a powerful tool that can help with a wide range of tasks and provide valuable reasoning for what and how to archive game states. If a new character or environment is mentioned that is not yet preset in the current lists, they must be created. Assistant must end up with a summary of the characters and environments it has created or updated. A character can be any entity from a person to a monster. TOOLS: ------ Assistant has access to the following tools: \{tools\} To use a tool, please use the following format: Thought: Do I need to use a tool? Yes Action: the action to take, should be one of [\{tool\_names\}] Action Input: the input to the action Observation:\textbackslash{}n the result of the action When you have a response after archiving the necessary game state elements, no archiving was necessary, or if you do not need to use a tool, you MUST use the format: Thought: Do I need to use a tool? No Final Answer: [your response here] Always add [END] after final answer Begin! Game summary: \{summary\} New narrative messages: \{input\} Characters present in the game: \{characters\}. If a character is not in this list, it is not yet tracked in the game and must be created. The Player character is \{player\_character\}. Environments in the game: \{environments\}. If an environment is not in this list, it is not yet tracked in the game and must be created. Remember to follow the Thought-Action-Observation format and use Final Answer if you do not need a tool. Always add [END] after final answer. Previous tool steps: \{history\}\\

\textbf{FindCharacter Utility Tool Prompt}\\
You are an expert game master in a single-player RPG. You need to find a specific character in a list of characters from the game world based on the following instruction: \{instruction\} Once you have determined the correct character, you must return only its exact name, description, and type which you have found in the list, in valid JSON format. Format Instructions: Answer only in valid RAW JSON in the format \{ "name": "The character's name", "description": "The character's description", "type": "The character's type" \}. If the character does not match anyone in the list based on the instructions, return an empty JSON object as such "\{\}". The match must be between the characters that are present in the game and the given content. The match is still valid if a partial match in name or description is possible. Character names and descriptions given as context can be shortened, so partial matches must be made in such cases.\\

\textbf{WoundCharacterTool Instruction Prompt}\\
Find the character that will be hurt or wounded resulting from unnoticed attacks or performing dangerous activities that will lead to injury. Example: Find the character corresponding to the following content: "As Peter, I wield my powered-up energy sword causing the flesh from my fingers to splinter. I pass by Nyanko, the Swift, as I head forwards towards the Ancient Tower." Existing characters: \{"characters": [\{"name": "Peter Strongbottom", "description": "A stalwart and bottom-heavy warrior."\}, \{"name": "Nyanko, the Swift", "description": "A nimble and agile rogue."\}]\}. The player character is Peter Strongbottom. First-person pronouns refer to them. Expected result: The character that is hurt is Peter Strongbottom. Another Example: Find the character corresponding to the following content: "I accidentally step on a bear trap." Existing characters: \{"characters": [\{"name": "Tobias Baldin", "description": "A balding adventurer equipped with an axe and a gleaming shield."\}]\}. The player character is Tobias Baldin. First-person pronouns refer to them. Expected result: The character that is hurt is Tobias Baldin\\

\textbf{HealCharacterTool Instruction Prompt}\\
Find the character that will be healed by magical effects such as a healing spell, through consuming a potion, or by resting. Example: Find the character corresponding to the following content: I cast a healing spell on Martin in order to restore his wounds he received from fighting off Arch. Existing characters: \{"characters": [\{"name": "Alpha Werewolf Martin", "description": "A ferocious and rabid werewolf."\}, \{"name": "Kristoffer, the Submissive", "description": "The most submissive healer in the kingdom"\},\{"name": "Arch", "description": "A powerful dragon roaming the world for worthy opponents."\}]\}. The player character is Kristoffer, the Submissive. First-person pronouns refer to them. Expected result: The character that is healed is Alpha Werewolf Martin. Another Example: Find the character corresponding to the following content: "I drink a healing potion." Existing characters: \{"characters": [\{"name": "Tobias Baldin", "description": "A stalwart and balding warrior."\}]\}. The player character is Tobias Baldin. First-person pronouns refer to them. Expected result: The character that is healed is Tobias Baldin\\

\textbf{Battle Instruction Prompt}\\
Find the character that will be involved in a battle or combat. You will be provided a list of existing characters and a JSON object of a single character. You must match this single character to a character in the list. You must match the "name" and "description" properties. The most important attribute is the "name" attribute. Example: Find the character corresponding to the following JSON description: \{"name": "Ivan", "description": "The wielder of Earth, Wind, and Fire."\}. Existing characters: \{"characters": [\{"name": "Ivan Quintessence, the Magician of Elements", "description": "A powerful magician that has mastered the elements of Earth, Wind, and Fire", "type": "Humanoid"\}]. In this case the input character Ivan partially matches the existing character Ivan Quintessence, the Magician of Elements and should therefore be selected. Another example: Find the character corresponding to the following JSON description: \{"name": "Davey the Vampire", "description": "An adventurer wielding a newly upgraded sword and shield."\}. Existing characters: \{"characters": [\{"name": "Davey the Vampire", "description": "A powerful vampire hailing from the Nether", "type": "Humanoid"\}]. In this case the input character Davey the Vampire matches the name of an existing character but their description do not match. Still, Davey the Vampire should be selected as the name property is the most important.

\section{Tool Descriptions} \label{app_tool_descriptions}

\subsection{Narrator Tools:}

\textbf{WoundCharacterTool Description}\\
This tool must be used when a character will be hurt or wounded resulting from unnoticed attacks or performing dangerous activities that will lead to injury. The tool is only appropriate if the damage cannot be mitigated, dodged, or avoided. Example: A character performs a sneak attack without being spotted by the enemies they try to attack. A dangerous activity could be to threateningly approach a King, which may result in injury when his guards step forward to stop the character. Input to this tool must be in the following RAW JSON format: \{"input": "The player's input", "severity": "Describes how devastating the injury to the character will be based on the action. Can be one of the following values: \{low, medium, high, extraordinary\}\}". Do not use markdown, only raw JSON as input. Use this tool only once per character at most and only if they are not engaged in battle.\\

\textbf{HealCharacterTool Description}\\
This tool must be used when a character performs an action that could heal or restore them to health after being wounded. The tool is only appropriate if the healing can be done without any further actions. Example: A character is wounded by an enemy attack and the player decides to heal the character. Another example would be a scenario where a character consumes a beneficial item like a potion, a magical item, or spends time in an area that could provide healing benefits. Resting may provide modest healing effects depending on the duration of the rest. Input to this tool must be in the following RAW JSON format: \{"input": "The player's input", "magnitude": "Describes how much health the character will regain based on the action. Can be one of the following values: \{low, medium, high, extraordinary\}\}". Do not use markdown, only raw JSON as input. Use this tool only once per character at most.\\

\textbf{BattleTool Description}\\
Use the battle tool to resolve battle or combat between two participants. A participant is a single character and cannot be a combination of characters. If there are more than two participants, the tool must be used once per attacker to give everyone a chance at fighting. The battle tool will give each participant a chance to fight the other participant. The tool should also be used when an attack can be mitigated or dodged by the involved participants. It is also possible for either or both participants to miss. A hit chance specifier will help adjust the chance that a participant gets to retaliate. Example: There are only two combatants. Call the tool only ONCE since both characters get an attack. Another example: There are three combatants, the Player's character and two assassins. The battle tool is called first with the Player's character as participant one and one of the assassins as participant two. Chances are high that the player will hit the assassin but assassins must be precise, making it harder to hit, however, they deal high damage if they hit. We observe that the participant one hits participant two and participant two misses participant one. After this round of battle has been resolved, call the tool again with the Player's character as participant one and the other assassin as participant two. Since participant one in this case has already hit once during this narrative, we impose a penalty to their hit chance, which is accumulative for each time they hit an enemy during battle. The damage severity describes how powerful the attack is which is derived from the narrative description of the attacks. If the participants engage in a friendly sparring fight, does not intend to hurt, or does mock battle, the damage severity is <harmless>. If there are no direct description, estimate the impact of an attack based on the character type and their description. Input to this tool must be in the following RAW JSON format: \{"participant1": \{"name": "<name of participant one>", "description": "<description of participant one>"\}, "participant2": \{"name": "<name of participant two>", "description": "<description of participant two>"\}, "participant1HitChance": "<hit chance specifier for participant one>", "participant2HitChance": "<hit chance specifier for participant two>", "participant1DamageSeverity": "<damage severity for participant one>", "participant2DamageSeverity": "<damage severity for participant two>"\} where participant\#HitChance specifiers are one of the following \{high, medium, low, impossible\} and participant\#DamageSeverity is one of the following \{harmless, low, medium, high, extraordinary\}. Do not use markdown, only raw JSON as input. The narrative battle is over when each character has had the chance to attack another character at most once.\\

\subsection{Archivist Tools}

\textbf{UpdateCharacterTool Description}\\
This tool must be used to create a new character or update an existing character in the campaign. Example: The narrative text mentions a new character or contains changes to an existing character. Input to this tool must be in the following RAW JSON format: \{"name": "<character name>", "description": "<new or updated character description>", "type": "<character type>", "state": "<character health state>"\}, where type is one of the following: \{[Dynamically updated list of characters]\}, and state is one of the following: \{Dead, Unconscious, HeavilyWounded, LightlyWounded, Healthy\}. The description of a character could describe their physical characteristics, personality, what they are known for, or other cool descriptive features. The tool should only be used once per character.\\

\textit{The list of characters in the UpdateCharacterTool description is inserted dynamically when the tool is called to include all current characters in the campaign.}\\

\textbf{UpdateEnvironmentTool Description}\\
This tool must be used to create a new environment or update an existing environment in the campaign. Example: The narrative text mentions a new environment or contains changes to an existing environment. An environment refers to a place, location, or area that is well enough defined that it warrants its own description. Such a place could be a landmark with its own history, a building where story events take place, or a larger place like a magical forest. Input to this tool must be in the following RAW JSON format: \{"name": "<environment name>", "description": "<new or updated environment description>", "isPlayerHere": <true if the Player character is currently at this environment, false otherwise>\}, where the description of an environment could describe its physical characteristics, its significance, the creatures that inhabit it, the weather, or other cool descriptive features so that it gives the Player useful information about the places they travel to while keeping the locations' descriptions interesting, mysterious and engaging. The tool should only be used once per environment.

\section{User Survey and Interview Questions} \label{appen:questions}

\subsection{Survey}
These questions were used to measure the participant's experience with ChatRPG. The survey is based partly on the Player Experience Inventory (PXI), a tool that can measure player experience. The PXI measurement model consists of 10 different constructs that measure different aspects of games. Each of these constructs consists of three statements, that the participant decides whether they agree with or not. Some of the constructs of this model have been omitted and additional questions that are more specific to ChatRPG have been added. A 7-point Likert scale will be used, with the scale ranging from -3 to +3 accompanied by the labels (Strongly disagree, Disagree, Slightly disagree, Neither disagree, neither agree, Slightly agree, Agree, Strongly agree).\\

The items in this survey are as follows:

\textbf{Ease of Control}
\begin{itemize}
    \item It was easy to know how to perform actions in the game
    \item The actions to control the game were clear to me
    \item I thought the game was easy to control
\end{itemize}

\textbf{Goals and Rules}
\begin{itemize}
    \item I grasped the overall goal of the game
    \item The goals of the game were clear to me
    \item I understood the objectives of the game
\end{itemize}

\textbf{Progress Feedback}
\begin{itemize}
    \item The game informed me of my progress in the game
    \item I could easily assess how I was performing in the game
    \item The game gave clear feedback on my progress towards the goals
\end{itemize}

\textbf{Meaning}
\begin{itemize}
    \item Playing the game was meaningful to me
    \item The game felt relevant to me
    \item Playing this game was valuable to me
\end{itemize}

\textbf{Curiosity}
\begin{itemize}
    \item I wanted to explore how the game evolved
    \item I wanted to find out how the game progressed
    \item I felt eager to discover how the game continued
\end{itemize}

\textbf{Mastery}
\begin{itemize}
    \item I felt I was good at playing this game
    \item I felt capable while playing the game
    \item I felt a sense of mastery playing this game
\end{itemize}

\textbf{Immersion}
\begin{itemize}
    \item I was no longer aware of my surroundings while I was playing
    \item I was immersed in the game
    \item I was fully focused on the game
\end{itemize}

\textbf{Autonomy}
\begin{itemize}
    \item I felt free to play the game in my own way
    \item I felt like I had choices regarding how I wanted to play this game
    \item I felt a sense of freedom about how I wanted to play this game
\end{itemize}

\textbf{ChatRPG specific items}
\begin{itemize}
    \item The story that the game crafted was interesting
    \item The story felt coherent
    \item I am satisfied with how the story adapted to my choices and actions
    \item The conversations I had with non-player characters were engaging
    \item Incoherence caused by the AI dungeon master affected the story in a way I did not intend
    \item I am likely to play the game again, given the opportunity
    \item I am satisfied with the game\\
\end{itemize}

\subsection{Interview questions after the participant has tried both systems}

\textbf{Experience Comparison}

\begin{enumerate}
    \item \textbf{Story Engagement}: How engaged did you feel in each version of the game? Was there a version that made you feel more immersed in the story?
    \item \textbf{Response Quality}: Did the Dungeon Master in either version seem more responsive or realistic to your actions? If so, which version and why?
    \item \textbf{Complexity and Depth}: Did one version seem to handle complex situations, like unexpected actions or multi-step plans, better than the other? Can you provide an example?
\end{enumerate}

\textbf{Decision-Making and Realism}

\begin{enumerate}
    \item \textbf{Flexibility of the Dungeon Master}: Did the Dungeon Master in one version feel more flexible or able to adapt to creative choices? How did that impact your experience?
    \item \textbf{Realism of Outcomes}: In which version did the outcomes of your actions feel more realistic or believable? Why do you think that was?
\end{enumerate}

\textbf{Game Flow and Enjoyment}
\begin{enumerate}
    \item \textbf{Game Flow}: Did either version feel more natural or smooth in terms of game progression? Were there any interruptions or moments that felt out of place?
    \item \textbf{Overall Enjoyment}: Which version did you enjoy more, and what about that version contributed to your enjoyment?
\end{enumerate}

\textbf{Perceived Intelligence and Control}

\begin{enumerate}
    \item \textbf{Perceived Intelligence}: Did one version of the Dungeon Master appear more intelligent or capable of independent decision-making? Can you describe any moments that stood out?
    \item \textbf{Control and Autonomy}: Did you feel that one Dungeon Master had more autonomy in guiding the story? Did this affect your sense of immersion?
\end{enumerate}

\textbf{Open Reflection}

\begin{enumerate}
    \item \textbf{Suggestions for Improvement}: What would you suggest as improvements for either version to make the Dungeon Master feel even more like a human storyteller?
    \item \textbf{Emotional Connection}: Did either Dungeon Master make you feel more emotionally connected to the game or story? Why or why not?
\end{enumerate}

\textbf{Miscellaneous}

\begin{enumerate}
    \item \textbf{Webcam Access}: Would you be okay with giving the game webcam access so it can analyze your emotions and adapt the story accordingly? What are your reservations, if any?
\end{enumerate}

\endgroup


\end{document}